\newif\ifnote 
\notefalse    

\newif\ifjhep 
\jheptrue    
\newif\ifprep 
\prepfalse    
\ifnote
\documentclass[11pt,a4paper,dvips]{scrartcl}
\usepackage{slashbox}
\usepackage{cite,mcite}
\usepackage{graphicx}
\usepackage{subfig}
\usepackage{lcd}
\usepackage{lcd_title,ifthen}
\usepackage{mathptmx}
\usepackage{helvet}
\usepackage{amsmath}
\usepackage{color}
\usepackage{units}
\usepackage{url}
\usepackage{caption}
\usepackage{exscale}
\usepackage{booktabs}
\usepackage[point]{rccol}
\usepackage[hmargin=2cm,vmargin=3.5cm]{geometry}
\makeatletter
\def\url@myurlfontstyle{%
  \@ifundefined{selectfont}{\def\UrlFont{\sf}}{\def\UrlFont{\small\ttfamily}}}
\makeatother
\long\def\symbolfootnote[#1]#2{\begingroup%
\def\thefootnote{\fnsymbol{footnote}}\footnote[#1]{#2}\endgroup}
 \lcdnote{2012-012}
\newlength{\capindent}
\newlength{\capwidth}
\newlength{\figwidth}
\newcommand{\icaption}[2][!*!,!]{\hspace*{\capindent}%
  \begin{minipage}{\capwidth}
    \ifthenelse{\equal{#1}{!*!,!}}%
      {\caption{#2}}%
      {\caption[#1]{#2}}
      \vspace*{3mm}
  \end{minipage}}
\fi
\ifjhep
\documentclass[11pt,a4paper]{article}
\usepackage{jheppub}
\usepackage{subfig}
\usepackage{ifthen}
\usepackage{units}
\usepackage[point]{rccol}
\fi
\def\susy#1{\ensuremath{\tilde{\mathrm{#1}}}}%

\def\chargino  #1{\ensuremath{\susy{\chi}_1^{#1}}}

\def\neutralino#1{\ensuremath{\susy{\chi}_{#1}^0}}
\def\ee{\ensuremath{e^+ e^-}}%
\def\pT{\ensuremath{p_T}}

\ifjhep
\title{
Physics performance for Scalar Electron, Scalar Muon and Scalar Neutrino searches at $\sqrt{s}=$ 3 TeV and 1.4 TeV
at CLIC
}
\author[1,2]{M. Battaglia,}
\author[3]{J-J. Blaising,}
\author[4]{J. S. Marshall,}
\author[1,3]{S. Poss,}
\author[4]{M. Thomson,}
\author[1]{A. Sailer,}
\author[1]{E. van der Kraaij}
\affiliation[1]{CERN CH-211 Geneva, Switzerland}
\affiliation[2]{Santa Cruz Institute of Particle Physics, University of California, Santa Cruz, CA 95064, USA}
\affiliation[3]{Laboratoire d'Annecy-le-Vieux de Physique des Particules, Annecy-le-Vieux, France}
\affiliation[4]{Cavendish Laboratory, University of Cambridge, Cambridge, United Kingdom}

\emailAdd{Jean-Jacques.Blaising@cern.ch}
\abstract{
The determination of scalar lepton and gaugino masses is
an important part of the programme of spectroscopic studies of Supersymmetry at a high energy
$e^+e^-$ linear collider.
In this article we present results of a study of the processes:
$e^+e^- \to \tilde e_R^+~\tilde e_R^- \to e^+e^- ~\neutralino{1}~\neutralino{1} $,
$e^+e^- \to \tilde \mu_R^+ ~\tilde \mu_R^- \to \mu^ + \mu^-~\neutralino{1}~\neutralino{1}$,
$e^+e^- \to \tilde e_L^+ ~\tilde e_L^- \to e^+~e^- ~\neutralino{2}~\neutralino{2} $ and
$e^+e^- \to \tilde \nu_e ~\tilde \nu_e \to e^+~e^- ~\chargino{+}~\chargino{-}$
in two Supersymmetric benchmark scenarios at $\sqrt{s}$\ =\ 3 TeV and 1.4 TeV at CLIC.
We characterize the detector performance, lepton energy resolution and boson mass resolution.
We report the accuracy of the production cross section measurements and
the $\tilde e_R,~ \tilde \mu_R,~
\tilde \nu_e,~ \chargino{\pm}$, and $\neutralino{1}$ mass determination,
estimate the systematic errors affecting the mass measurement 
and discuss the requirements 
on the detector time stamping capability and beam polarization.
The analysis accounts for the CLIC beam energy spectrum and the dominant beam-induced background.
The detector performances are incorporated by full simulation and reconstruction of the events
within the framework of the  CLIC\_ILD\_CDR detector concept.

}
\fi

\begin{document}
\message{**:main 11}

\ifjhep
\message{**:jhep maketitle}
\date{\currenttime}
\maketitle
\flushbottom
\fi
%
%
\section{Introduction}
One of the main objectives of linear collider experiments is the precision spectroscopy
of new particles predicted in theories of physics beyond the Standard Model (SM), such as
Supersymmetry (SUSY).
In this article, we study the production of the supersymmetric partners of the muon, electron and
neutrino in two specific SUSY benchmark points, where we assume R--parity conservation within the so-called
constrained Minimal Supersymmetric extension of the SM (cMSSM).
In this model the neutralino (\neutralino{1}) is the lightest supersymmetric particle. 
Table~\ref{tab:params} shows the masses and the branching ratios of the supersymmetric particles 
for the two benchmark points P1 and P2. 
\begin{table} [htbp]               
\centering
\caption{Benchmark parameters of the considered SUSY model. }
\label{tab:params}
\begin{tabular}{ l c c l }
\hline
  Benchmark point & P1 ($\sqrt{s}=$ 3 TeV) & P2 ($\sqrt{s}=$ 1.4 TeV) & \\ \hline
  $\neutralino{1}$ mass & 340 & 357 & GeV \\ 
  $\chargino{\pm}, \neutralino{2}$ mass & 643, 917 & 487, 911 & GeV \\ 
  $\tilde e_R^{\pm}, \tilde \mu_R^{\pm}$ mass & 1011, 1011 & 559, 559 & GeV \\
  $\tilde e_L^{\pm}, \nu_e$ mass & 1110, 1097 & 650, 644 & GeV \\ 
  Br ($\tilde \ell{_R^\pm} \to \ell{^\pm} ~\neutralino{1}$)  & 100 & 100 & \% \\ 
  Br ($\tilde e_L \to e^- ~\neutralino{1}$  & 16 & 19 & \% \\ 
  Br ($\tilde e_L \to e^- ~\neutralino{2}$) & 29 & 28 & \% \\
  Br ($\tilde \nu_e \to e^-~\chargino{+}~$) & 56 & 53 & \%\\ \hline
\end{tabular}
\end{table}

\begin{figure}[htbp]
\centering
\resizebox{\textwidth}{!} {
\begin{tabular}{c}
\includegraphics[width=1.0\textwidth]{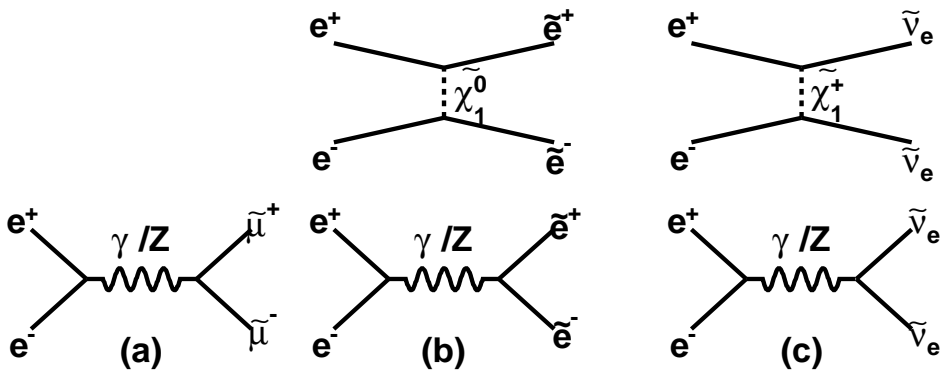}
\end{tabular}
}
\caption{Slepton production diagrams: scalar muons (a), scalar electrons (b),
and scalar neutrinos (c).}
 \label{fig:diagrams}
\end{figure}

For both benchmark points the Higgs boson mass is 120 GeV.
Smuons are produced in pairs through \mbox{$s$-channel} $\gamma/\mathrm{Z}$ exchange,
selectrons and sneutrinos are pair produced through \mbox{$s$-channel} $\gamma/\mathrm{Z}$ exchange
or \mbox{$t$-channel} \neutralino{1} and \chargino{\pm} exchange respectively,
see Figure~\ref{fig:diagrams}.

The cross sections, the decay modes, and the cross sections times the branching
ratio of the signal processes are given in Table~\ref{tab:signal}. 
In the processes $\ee \rightarrow \tilde \ell{_R^+} \tilde \ell{_R^-}$ each $\tilde \ell{_R^{\pm}}$
decays into an ordinary lepton and a \neutralino{1}; the \neutralino{1} is stable and escapes
detection due to its weakly interacting nature.
Therefore, the experimental signature of these processes is two oppositely
charged leptons plus missing energy.
For the processes \mbox{$e^+e^- \to \tilde e_L^ + \tilde e_L^- \to e^+~e^- \neutralino{2}~\neutralino{2} $}
and
\mbox{$e^+ e^- \to \tilde\nu_e \tilde\nu_e \to e^+ e^- \chargino{+} \chargino{-}$} 
the signature is an \ee pair, four jets, and missing energy.
Measuring the lepton energy distributions of these four processes allows the
determination of their production cross sections and of the
$\tilde e_R,~ \tilde \mu_R,~\tilde \nu_e, ~\chargino{\pm}$, and $\neutralino{1}$ masses.
The aim of this study is to:
\begin {itemize}
\item Characterize the detector performance, namely lepton energy resolution, and boson mass resolution.
\item Assess the statistical accuracy of the cross section measurements and the mass determination.
\item Estimate the systematic errors, affecting the mass measurements, 
related to the event selection and the luminosity spectrum knowledge.
\item Set the requirements for the detector time stamping capability
and beam polarization.     
\end {itemize}
The results presented in this article improve and supersede the previous results \cite{LCWS11}
obtained at 3 TeV only.
\begin{table} [tp]
  \centering
\caption{Signal processes, decay modes, cross sections, and
cross sections times branching ratio ($\sigma \times Br$) at  $\sqrt{s}=$ 3 TeV and \mbox{1.4 TeV}.}
\label{tab:signal}
\resizebox{\textwidth}{!} {
    \begin{tabular}{l l c c c c}
\hline
$\sqrt{s}$ (TeV)                                               &           &3.0        &3.0             &1.4     &1.4  \\
Process                                                  &Decay Mode &$\sigma$  &$\sigma \times Br$ &$\sigma$  &$\sigma \times Br$ \\
                                                         &           &   (fb)        &   (fb)             &   (fb)     &   (fb) \\
\hline
$\ee \rightarrow \tilde \mu_R^+ \tilde \mu_R^-$     &$\mu^+\mu^- \neutralino{1} \neutralino{1}$ &~0.70  &~0.70 &~1.53   &~1.53 \\
$\ee \rightarrow \tilde e_R^+ \tilde e_R^-$         &$ e^+e^- \neutralino{1} \neutralino{1}$   &~6.10  &~6.10  &~5.91 &~5.91 \\
$\ee \rightarrow \tilde e_L^+ \tilde e_L^-$         &$ e^+e^- \neutralino{2} \neutralino{2}
~\rightarrow e^+e^- H/Z^0  H/Z^0 \neutralino{1} \neutralino{1}$  &~3.06 &~0.26 &~0.73 &~0.06 \\
$\ee \rightarrow \tilde \nu_e \tilde \nu_e $    &$ e^+  e^- \chargino{\pm} \chargino{\pm}
\rightarrow e^+e^- W^+  W^- \neutralino{1} \neutralino{1}$ &~13.7 &~4.30  &~5.37 &~1.51\\
\hline
\end{tabular}
}
\end{table}

\section{Event Simulation}
SUSY signal events and SM background events are generated using
the WHIZARD program~\cite{Whizard:2008},
assuming zero polarisation of the electron and positron beams.
WHIZARD is interfaced to {\sc Pythia~6.4}~\cite{Sjostrand:2006za} for fragmentation and hadronization.
For the generation of processes involving supersymmetric particles,
the SUSY parameters are entered into WHIZARD using the Les Houches format~\cite{Alwall:2006yp}.
The physics backgrounds simulated for this study are listed
in Table~\ref{tab:background}.
\rcRoundingfalse
\begin{table} [tp]
\centering
\caption{Background processes, decay modes and
cross sections times branching ratio, $\sigma \times Br$, without and with preselection cuts, at 3 TeV and \mbox{1.4 TeV}. }
\label{tab:background} 
\resizebox{\textwidth}{!} {
\begin{tabular}{ l l R{4}{2} R{2}{3} R{5}{2} R{2}{4} }
\hline
\multicolumn{1}{c}{$\sqrt{s}$ (TeV)} &  \multicolumn{1}{c}{~} & \multicolumn{1}{c}{~3.0}  & \multicolumn{1}{c}{~3.0}  & \multicolumn{1}{c}{~1.4}  & 
\multicolumn{1}{c}{~1.4} \\
\multicolumn{1}{c}{Generator cuts} &  \multicolumn{1}{c}{~} & \multicolumn{1}{c}{~no}  & \multicolumn{1}{c}{~yes}  & \multicolumn{1}{c}{~no}  & 
\multicolumn{1}{c}{~yes} \\
\multicolumn{1}{c}{Process} &  \multicolumn{1}{c}{Decay mode} & \multicolumn{1}{c}{~$\sigma\times Br$}  & \multicolumn{1}{c}{~$\sigma\times Br$ }  & 
\multicolumn{1}{c}{~$\sigma\times Br$ } & \multicolumn{1}{c}{~$\sigma\times Br$ } \\
\multicolumn{1}{c}{~} &  \multicolumn{1}{c}{~} & \multicolumn{1}{c}{~fb}  & \multicolumn{1}{c}{~fb}  & \multicolumn{1}{c}{~fb}  & 
\multicolumn{1}{c}{~fb} \\
\hline
$\ee \rightarrow \mu^+ \mu^-        $ & $\mu^+ \mu^-$               &81.9  &0.65 &147.5 &0.72  \\
$\ee \rightarrow \mu^+ \nu_{e} \mu^- \nu_{e}$ & $\mu^+ \mu^-    $   &65.6  &3.5 &44.7 &2.12\\
$\ee \rightarrow \mu^+ \nu_{\mu} \mu^- \nu_{\mu}$ & $\mu^+ \mu^-$   &6.2  &2.2 &14.6 &5.73 \\
$\ee \rightarrow \mu^+ \mu^- e^+ e^-$ & $\mu^+ \mu^-$               &1689.1  &41.54 &1608.0 &23.8 \\
$\ee \rightarrow \mathrm{W^+ \nu W^- \nu}$ & $ \mu^+ \mu^-    $     &92.6  &2.4 &29.5 &0.73  \\
$\ee \rightarrow \mathrm{Z^0 \nu Z^0 \nu}$ & $ \mu^+ \mu^-    $     &40.5  &0.002 &10.8 & 0.0007  \\
$\ee \rightarrow \mathrm{All~SUSY}-(\tilde \mu_R^+ \tilde \mu_R^-) $ & $ \mu^+ \mu^- $   &0.31   &0.31 &0.12 &0.12 \\
\hline
$\ee \rightarrow e^+ e^-                 $ & $e^+ e^-         $     &6226.0  &77.1  &21180. &90.6\\
$\ee \rightarrow e^+ \nu_{e} e^- \nu_{e} $ & $e^+ e^-         $     &179.3   &91.1  &200.8 &96.4 \\
$\ee \rightarrow W^+ \nu W^- \nu$ & $ e^+ e^-        $              &92.6    &2.4   &29.5 &0.73 \\
$\ee \rightarrow Z^0 \nu Z^0 \nu$ & $ e^+ e^-        $              &40.5    &0.002 &10.8 &0.0007\\
$\ee \rightarrow \mathrm{All~SUSY}-(\tilde e_R^+ \tilde e_R^-) $ & $ e^+ e^- $     &1.04   &1.04 &1.77 &1.77  
\\
\hline
$\ee \rightarrow W^+ W^- Z^0    $ & $ e^+ e^- W^+ W^- $             &1.4  &0.61  &1.84 &0.84\\
$\ee \rightarrow Z^0 Z^0 Z^0    $ & $ e^+ e^- Z^0 Z^0 $             &0.5  &0.023 &0.75 &0.038\\
$\ee \rightarrow \mathrm{All~SUSY}-(\tilde e_L^+ \tilde e_L^-$, $\tilde \nu_e \tilde \nu_e) $ &$e^+e^-
WW/HH/Z^0Z^0 $   &0.77  &0.12 &0.67 &0.10\\
\hline
\end{tabular}
}
\end{table}

\begin{figure}[tp]
\centering
\resizebox{\textwidth}{!} {
\begin{tabular}{c}
\subfloat[ $\ee \rightarrow  \tilde \mu_R^+ \tilde \mu_R^-,~ \sqrt{s}=$ 3 TeV ]
{\includegraphics[width=0.50\textwidth,clip]{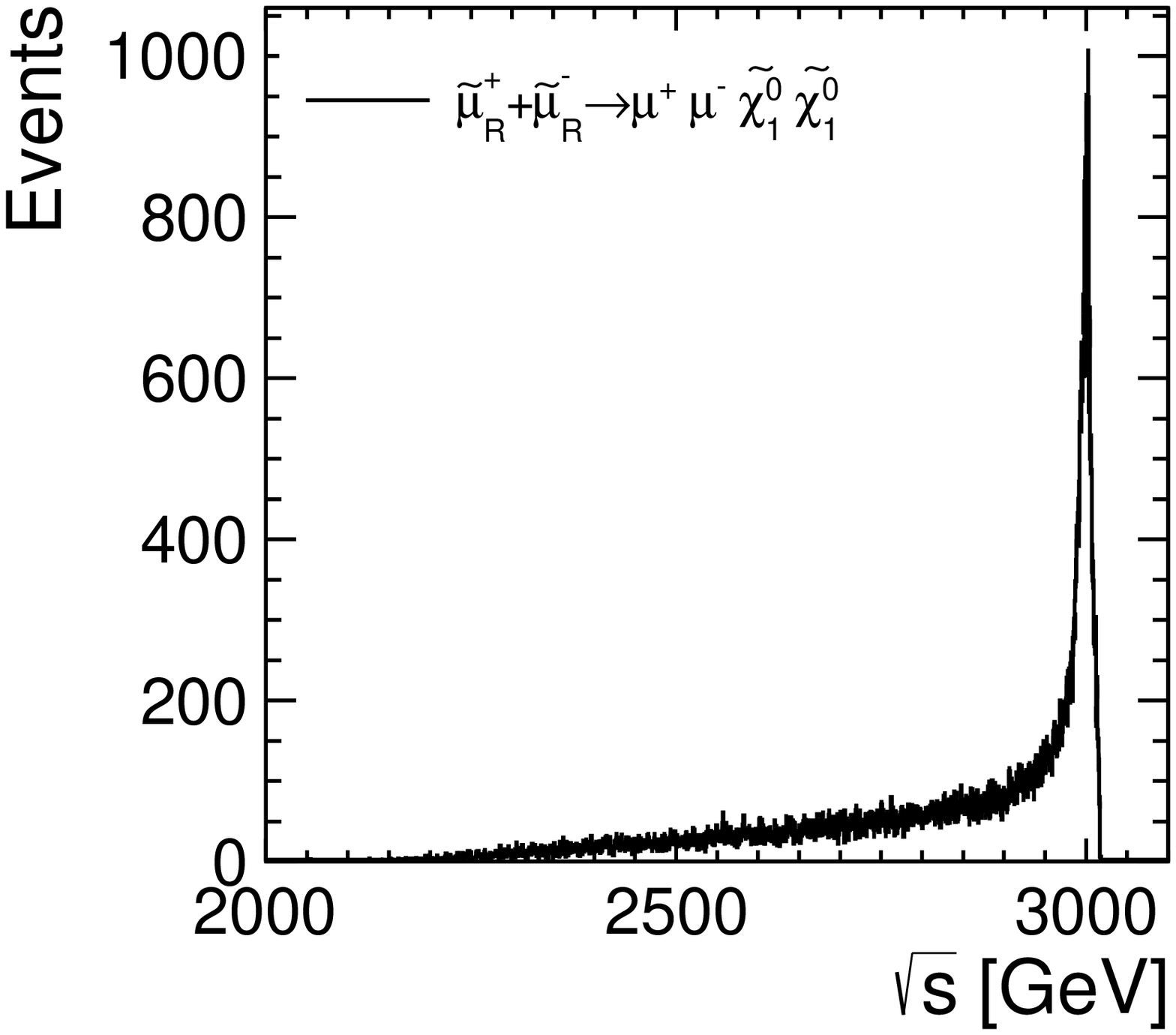}}
\subfloat[$\ee \rightarrow   \tilde e_R^+ \tilde e_R^-,~ \sqrt{s}=$ \mbox{1.4 TeV}]
{\includegraphics[width=0.50\textwidth,clip]{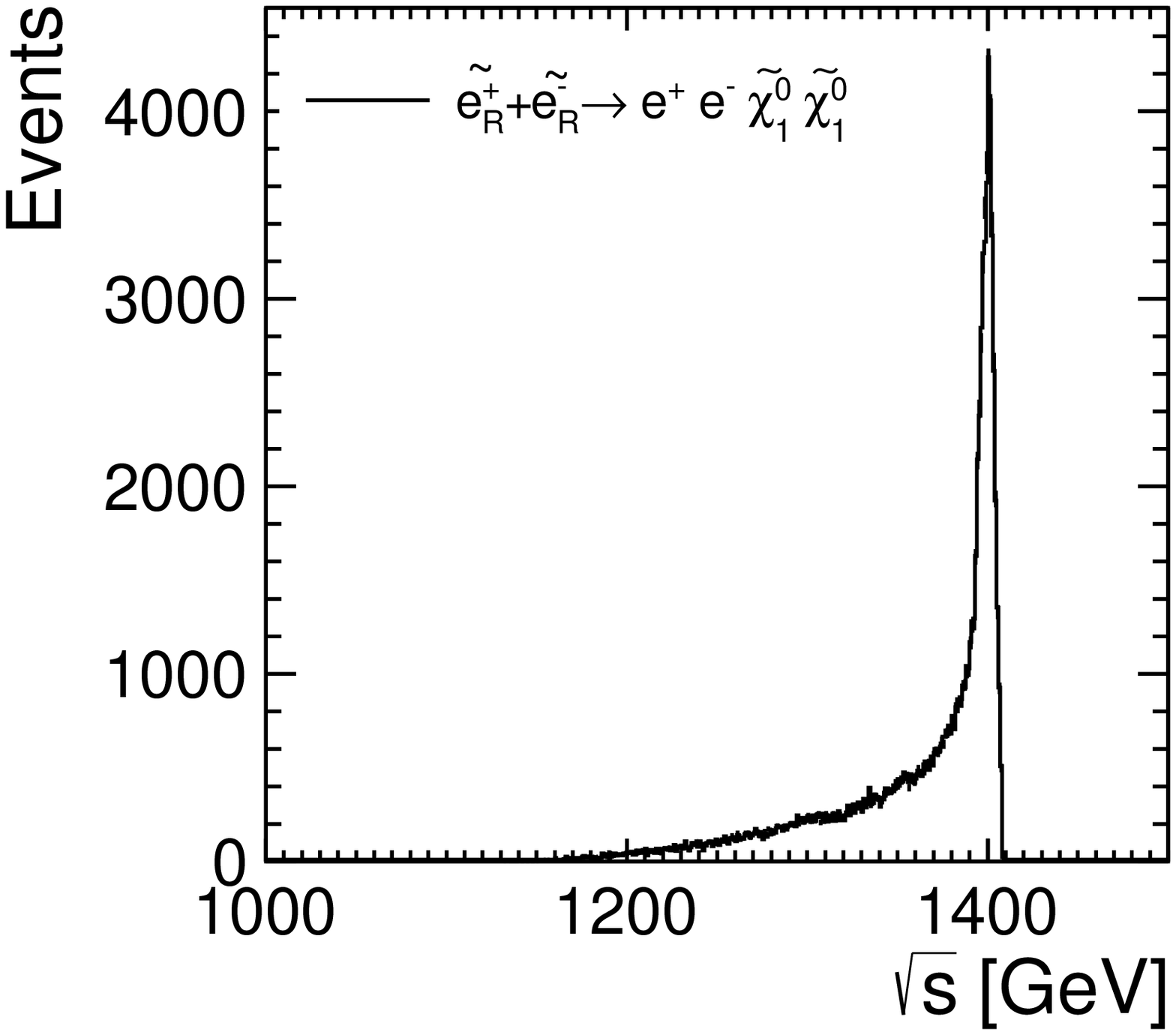}}
\end{tabular}
}
\caption{Centre-of-mass energy spectrum for the processes:
$\ee \rightarrow  \tilde \mu_R^+ \tilde \mu_R^-,~ \sqrt{s}=$ 3 TeV (a), and
\mbox{$\ee \rightarrow  \tilde e_R^+ \tilde e_R^-~, \sqrt{s}=$ 1.4 TeV (b)}.
}                
\label{fig:H1ECM}
\end{figure}

Beamstrahlung effects on the luminosity spectrum are included using results of the CLIC beam simulation
for the CDR accelerator parameters~\cite{Braun:2008zzb}.
There are three sources of the centre-of-mass energy spread: the momentum spread in the linac,
the beamstrahlung which creates a long tail, and initial state radiation (ISR).
The first two 
are collectively refererred to as ``luminosity spectrum''.
The luminosity spectrum is obtained from the
{\sc GuineaPig}~\cite{c:thesis} beam simulation; it is used as input
to WHIZARD in which initial state radiation and final state radiation (FSR) are enabled.
Figure~\ref{fig:H1ECM} shows the $\sqrt{s}$ distributions
for the processes $\ee \rightarrow \tilde \mu{_R^+} \tilde \mu{_R^-}$, at $\sqrt{s}=$ \mbox{3 TeV}
and $\ee \rightarrow  \tilde e_R^+ \tilde e_R^-~,$ at $\sqrt{s}=$ \mbox{1.4 TeV}.
Integrated luminosities of 2000~fb$^{-1}$ and 1500~fb$^{-1}$ are assumed at 3.0 and \mbox{1.4 TeV} respectively.
At  $\sqrt{s}=$ \mbox{3 TeV} an integrated luminosity of $\mathrm {2000~fb^{-1} }$ corresponds
to $\simeq$4 years (1 year = $10^{7}$~s) of run at the nominal CLIC luminosity of
5.9$\times$10$^{34}$~cm$^{-2}$s$^{-1}$.
At \mbox{1.4 TeV} the nominal luminosity is 3.2$\times$10$^{34}$~cm$^{-2}$s$^{-1}$.

The physics background cross sections
of the $\ee \rightarrow \tilde \ell{_R^+} \tilde \ell{_R^-}$
processes are very large, see Table~\ref{tab:background}. Taking into account the
luminosity assumptions, the simulation and
reconstruction of the background events would require very large computing and storage resources.
To optimize the use of these resources preselection cuts are applied
after generation of the background events. The preselection requires two opposite 
charged leptons ($L1$ and $L2$) and the following conditions:
%
\begin{itemize}
\item $\pT (L1 \mathrm{~and~} L2) > $ 4~GeV and $ 10^\circ < \theta (L1 \mathrm{~and~} L2) < 170^\circ $
\item  $4^\circ < \Delta \phi (L1,L2) < 176^\circ $, $\pT (L1,L2) > $ 10 GeV
and $M (L1,L2) > $ 100 GeV
\end{itemize}
where $\pT $ is the transverse momentum, $\theta$ the polar angle of the lepton,
$\Delta \phi(L1,L2)$ the acoplanarity of the leptons,
$\pT \mathrm{(L1,L2)}$ the vector sum of the $\pT$ of the two leptons, and
$M \mathrm{(L1,L2) }$ the invariant mass of the two leptons.
Table~\ref{tab:background} shows the decay modes, and the cross section times branching ratio values
without and with preselection cuts.
For  the signal samples, these cuts are also applied after full simulation and reconstruction.
The simulation is performed using the {\sc Geant4}-based~\cite{Agostinelli:2002hh}
{\sc Mokka} program~\cite{MoradeFreitas:2004sq} with the CLIC\_ILD\_CDR
detector geometry \cite{geom:2011},
which is based on the ILD detector concept \cite{loi:2009} being developed for the ILC.
%

\section{Event Reconstruction}
Events are subsequently reconstructed using the {\sc Marlin} reconstruction
program~\cite{Gaede:2006pj}.
The tracking systems of the CLIC detectors are designed to provide excellent momentum
measurement for charged particle tracks.
The track momenta and calorimeter data are input to the {\sc PandoraPFA} algorithm~\cite{Marshall:2010, LCD:2011-028}
which
performs particle flow (PFO) reconstruction, including particle identification and returns the best estimate for the momentum 
and energy of the particles.
\subsection{Two Lepton final states}
The energy of the lepton is reconstructed from the momentum of the charged
particle track and corrected for final state radiation and  bremsstrahlung.
The energy of photons and $\ee$ pairs from conversions within a cone of $20^\circ$
around the reconstructed lepton direction is added to the energy from the track.
Figure~\ref{fig:H1LEA_BX000_2} 
shows, for the process $\ee \rightarrow \tilde e{_R^+} \tilde e{_R^-}$, 
the true and reconstructed lepton energy distributions
without (a) and with (b) photon radiation correction.
For the process $\ee \rightarrow \tilde \mu{_R^+} \tilde \mu{_R^-}$
the photon radiation corrections are much smaller.
For both processes there is a good agreement between the true and reconstructed lepton
energy distributions when photon radiation corrections are applied.

Table~\ref{tab:RecEffi} shows the
reconstruction efficiencies, $\mathrm{ \epsilon_R}$, for the signal processes.
For the process $\ee \rightarrow \tilde \ell{_R^+} \tilde \ell{_R^-}$,
$\mathrm{ \epsilon_R}$ is the number of
good reconstructed lepton pairs divided by the number of generated lepton pairs.
A lepton is considered as good when the reconstructed lepton matches the generated particle in space within $2^\circ$. 
For the process $\ee \rightarrow \tilde \mu{_R^+} \tilde \mu{_R^-}$, at 3 TeV and \mbox{1.4 TeV}, there is an inefficiency of
about 2.5\%; 2.0\% is due to the cut on the lepton angle and 0.5\% is coming from muon
misidentification.
For the process $\ee \rightarrow \tilde e{_R^+} \tilde e{_R^-}$, at 3 TeV, there is an inefficiency of
6.5\%; 4.0\% is due to the cut on the lepton angle and 2.5\% is coming from electron reconstruction or misidentification.
At \mbox{1.4 TeV} the inefficiency is 5.5\%;
3.0\% is due to the cut on the lepton angle and 2.5\% is coming from electron reconstruction or misidentification.

\begin{table} [tp]
\centering
\caption{
Reconstruction efficiency, $\epsilon_R$, without and with $\gamma\gamma \to$~hadrons overlaid
for the different signal processes, at  $\sqrt{s}=$ 3 TeV and \mbox{1.4 TeV}. The statistical error on these efficiencies is 
$\sim 0.5\%$. }	
\label{tab:RecEffi}
\resizebox{\textwidth}{!} {
\begin{tabular}{ l l c c c c }
\hline
$\sqrt{s}$ (TeV)   &                      & 3 & 3 & 1.4 & 1.4 \\ 
Process &Decay Mode & $\epsilon_R$ &$\epsilon_R$  &$\epsilon_R$  &$\epsilon_R$ \\
        &             &without $\gamma \gamma$         &with $\gamma \gamma$    &without $\gamma \gamma$   &with $\gamma \gamma$ \\
\hline
$\ee \rightarrow  \tilde \mu_R^+ \tilde \mu_R^- $ &$\mu^+  \mu^- \neutralino{1} \neutralino{1}$
&~~0.975 &~~0.965 &~~0.975 &~~0.975 \\
$\ee \rightarrow  \tilde e_R^+ \tilde e_R^- $ &$e^+e^- \neutralino{1} \neutralino{1}$
&~~0.935 &~~0.905&~~0.944 &~~0.930\\
$\ee \rightarrow  \tilde e_L^+ \tilde e_L^- $
&$e^+e^- \neutralino{2} \neutralino{2} \rightarrow e^+  e^- H/Z^0  H/Z^0 \neutralino{1} \neutralino{1}$
&~~0.66~ &~~0.63~ &~~0.61~  &~~0.57~\\
$\ee \rightarrow \tilde\nu_e \tilde\nu_e$ &$e^+e^- \chargino{\pm} \chargino{\pm}
\rightarrow e^+ e^- W^+  W^- \neutralino{1} \neutralino{1}$
&~~0.49~ &~~0.46~ &~~0.43~ &~~0.40~\\
\hline
\end{tabular}
}
\end{table}

The energy resolution is characterized using:
$\Delta E/E_{\mathrm{True}}^2$, where
\mbox{$ \Delta E = E_{\mathrm{True}} - E_{\mathrm{Reco}} $},
$E_{\mathrm{True}}$ is the lepton energy at generator level before final state radiation or bremsstrahlung,
and $E_{\mathrm{Reco}}$ is the reconstructed lepton energy with photon radiation corrections.

Figure~\ref{fig:H1LEAT2_H1L2A_BX000} (a) and (b) show the lepton energy resolution, for the 
two lepton final state processes at $\sqrt{s}=$ 3 TeV
and without beam induced background, $\gamma \gamma \to$ hadrons.
The resolution is parametrised using the sum of two Gaussian functions G1 and G2; G1 for the peak
and G2 for the tails.
For the muons the r.m.s. of G1 is 
$\unit[1.5 \cdot 10^{-5}]{\text{GeV}^{-1}}$,
and the r.m.s. of G2 is $\unit[4.9 \cdot 10^{-5}]{\text{GeV}^{-1}}$.
Only 4.1\% of the events are outside of the central region; 
the central region of the distribution is defined within the interval
$\Delta E / E_{\text{True}}^2 = \pm 0.5 \cdot \unit[10^{-3}]{\text{GeV}^{-1}}$. 
The electron energy resolution is described by the Gaussian G1 with a very similar r.m.s. as that for
muons, 
$ \unit[1.4 \cdot 10^{-5}]{\text{GeV}^{-1}}$,
however, even with bremsstrahlung recovery, about 30\% of the events are outside the central region.
These are due to cases where final state radiation and bremsstrahlung are not sufficiently well accounted for;
the tails are reasonably well described by the Gaussian G2 with 
\mbox{r.m.s. = $\unit[7.7 \cdot 10^{-5}]{\text{GeV}^{-1}}$}.

\subsection{Two leptons and four jets final states}
For the processes $\ee \rightarrow  \tilde e_L^+ \tilde e_L^- $ and $\ee \rightarrow  \tilde \nu_e \tilde
\nu_e $, the parton topology signature required is two leptons and four quarks.
After the reconstruction of all the particles in the event, the jet finder
program {\sc FastJet} ~\cite{Fastjet:2010} is used to reconstruct jets.
The jet algorithm used is the inclusive anti-kt method~\cite{LCD:2010-006}; 
The choice of cylindrical coordinates is optimal since the $\gamma \gamma \to$ hadrons
events are forward boosted, similarly to the underlying events in pp
collisions for which the anti-kt clustering has been optimised.
The R parameter cut value is 1 and the minimum jet energy required is 20
GeV at $\sqrt{s}=$ 3 TeV and 10 GeV at \mbox{1.4 TeV}.
An event is retained if six jets are found and if two of the jets are identified as isolated leptons.
Table~\ref{tab:RecEffi} shows the reconstruction efficiencies of both processes,
$\mathrm{ \epsilon_R}$ is the number of reconstructed six jet events, with two leptons,
divided by the number of generated events with two leptons and four quarks.

\begin{figure}[htbp]
\centering
\resizebox{\textwidth}{!} {
\begin{tabular}{c}
 \subfloat[without photon energy correction.]
{\includegraphics[width=0.50\textwidth,clip]{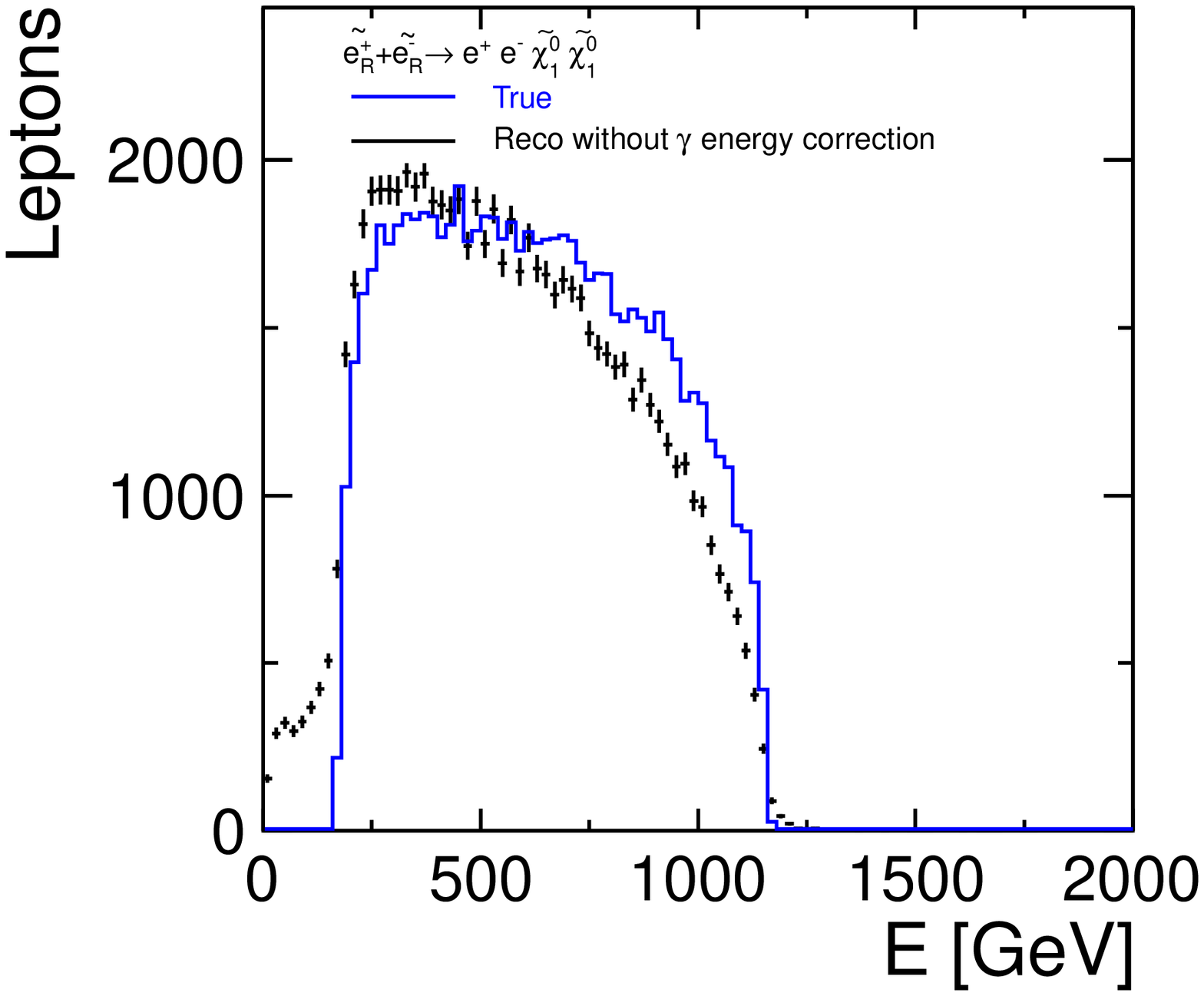}}
 \subfloat[with photon energy correction.]
{\includegraphics[width=0.50\textwidth,clip]{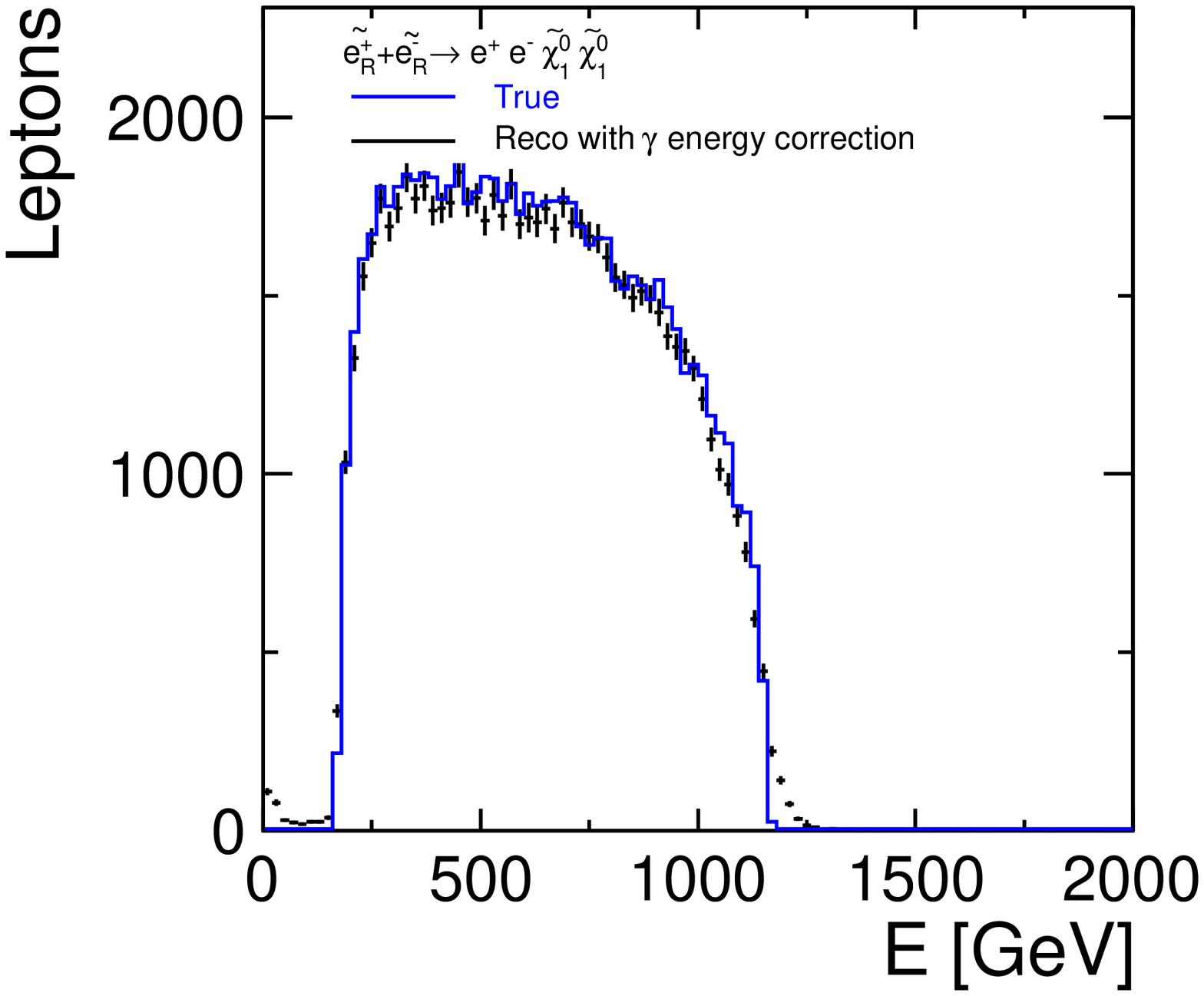}}
\end{tabular}
}
\caption{
Process $e^+e^- \rightarrow \tilde e_R^+ \tilde e_R^-$ at  $\sqrt{s}=$ 3 TeV:
true and reconstructed electron energy distributions, without (a), and with photon energy correction(b).}
\label{fig:H1LEA_BX000_2}
%
\centering
\resizebox{\textwidth}{!} {
\begin{tabular}{c}
 \subfloat[ $e^+e^- \rightarrow \tilde \nu_e^+ \tilde \nu_e^-$]
{\includegraphics[width=0.45\textwidth,clip]{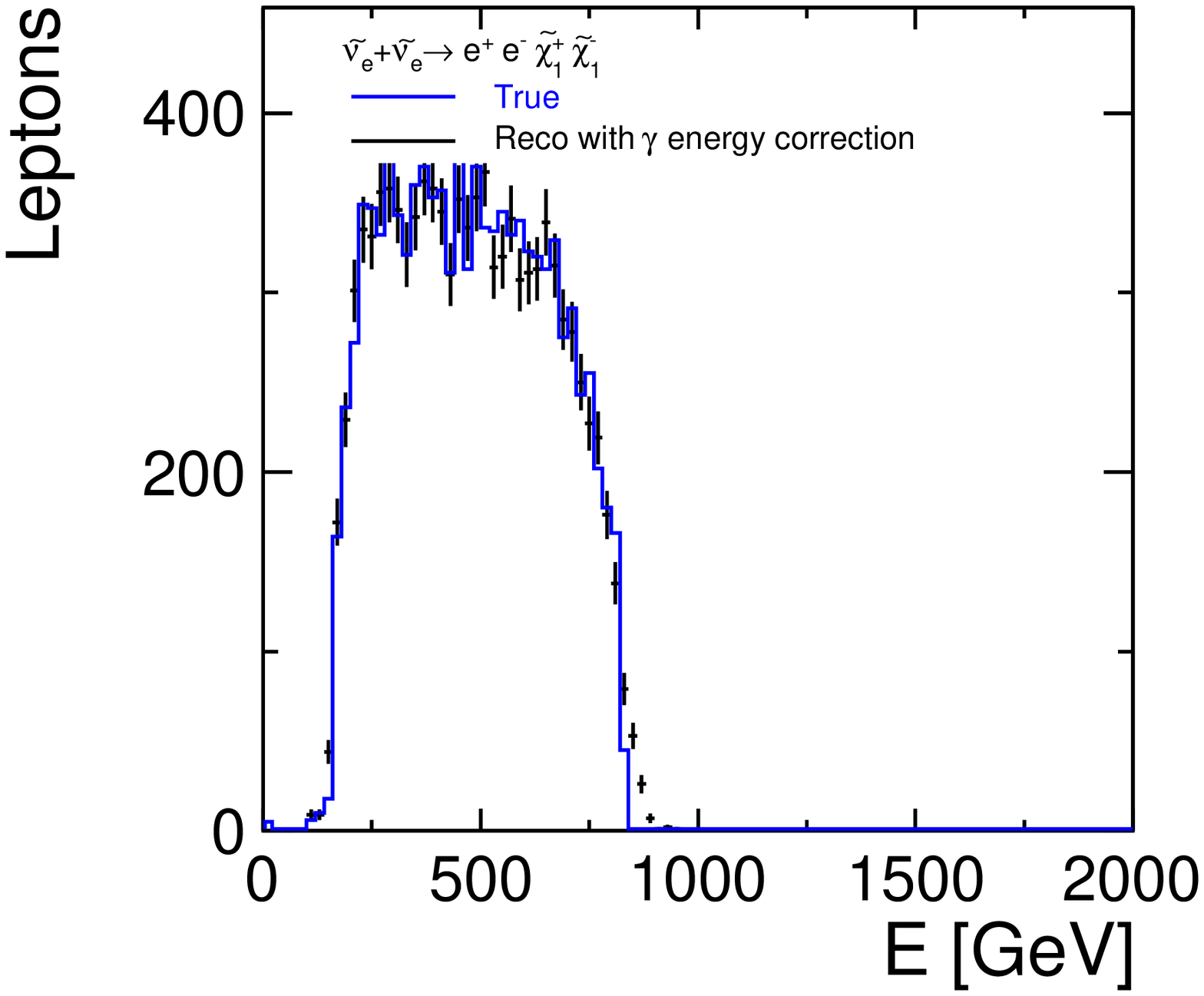}}
 \subfloat[ $e^+e^- \rightarrow \tilde e_L^+ \tilde e_L^-$]
{\includegraphics[width=0.45\textwidth,clip]{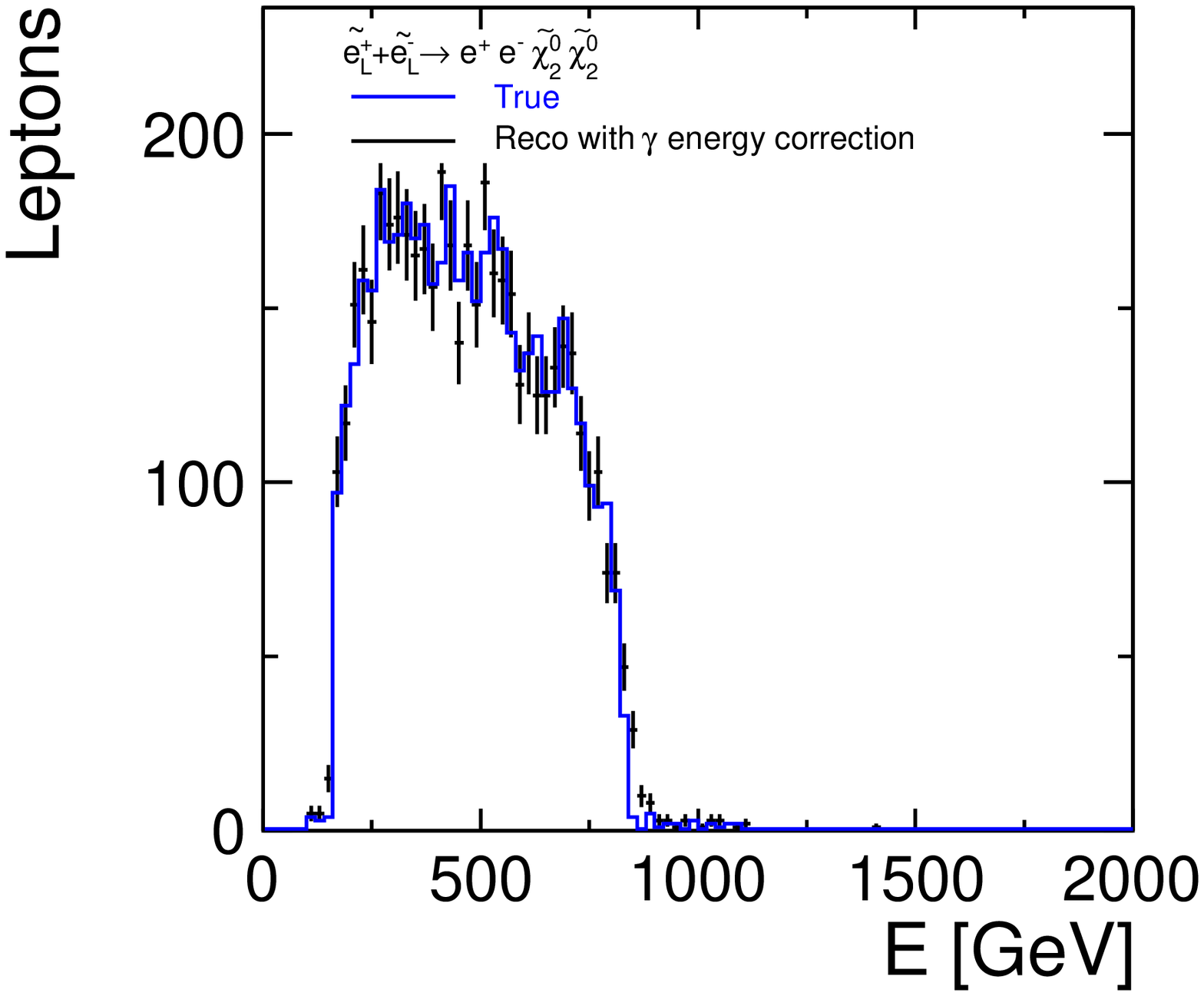}}
\end{tabular}
}
\caption{                
True and reconstructed electron energy distribution, with photon energy correction, for processes
$e^+e^- \rightarrow \tilde \nu_e^+ \tilde \nu_e^-$ (a), and
$e^+e^- \rightarrow \tilde e_L^+ \tilde e_L^-$ (b) at $\sqrt{s}=$ 3 TeV.}
\label{fig:H1LEA_BX000_3}
\end{figure}

Figure~\ref{fig:H1LEA_BX000_3} shows the electron energy distribution
for the processes $\ee \rightarrow  \tilde \nu_e \tilde \nu_e $ (a) and
$\ee \rightarrow  \tilde e_L^+ \tilde e_L^- $ (b).
There is good agreement between the true and reconstructed electron
energy distributions when photon radiation corrections are applied.

\begin{figure}[htbp]
\centering
\resizebox{\textwidth}{!} {
\begin{tabular}{c}
\hspace{-1.cm}
\subfloat[$\ee \rightarrow  \tilde \mu_R^+ \tilde \mu_R^- $]
{\includegraphics[width=0.50\textwidth,clip]{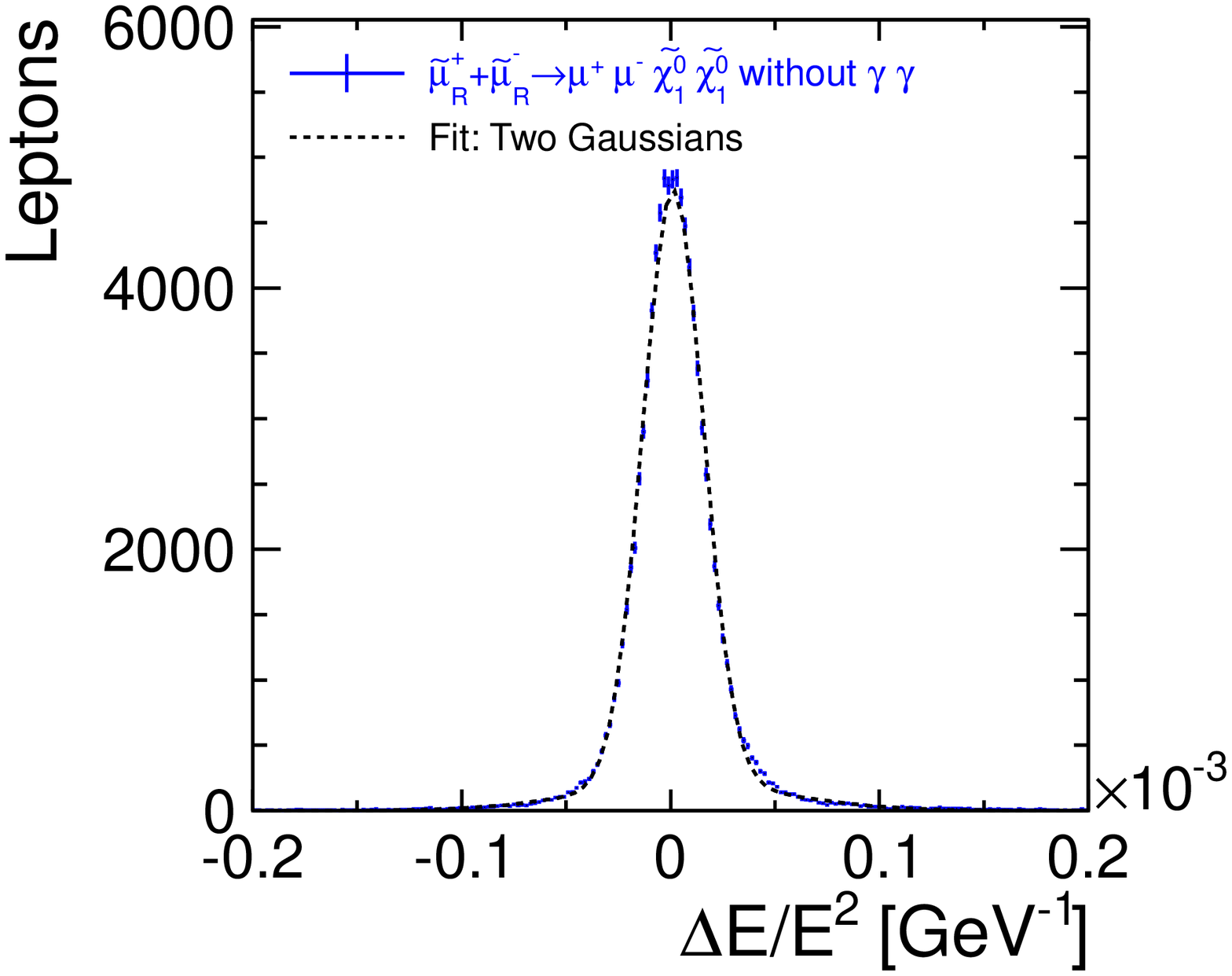}}
\subfloat[$\ee \rightarrow  \tilde e_R^+ \tilde e_R^- $]
{\includegraphics[width=0.50\textwidth,clip]{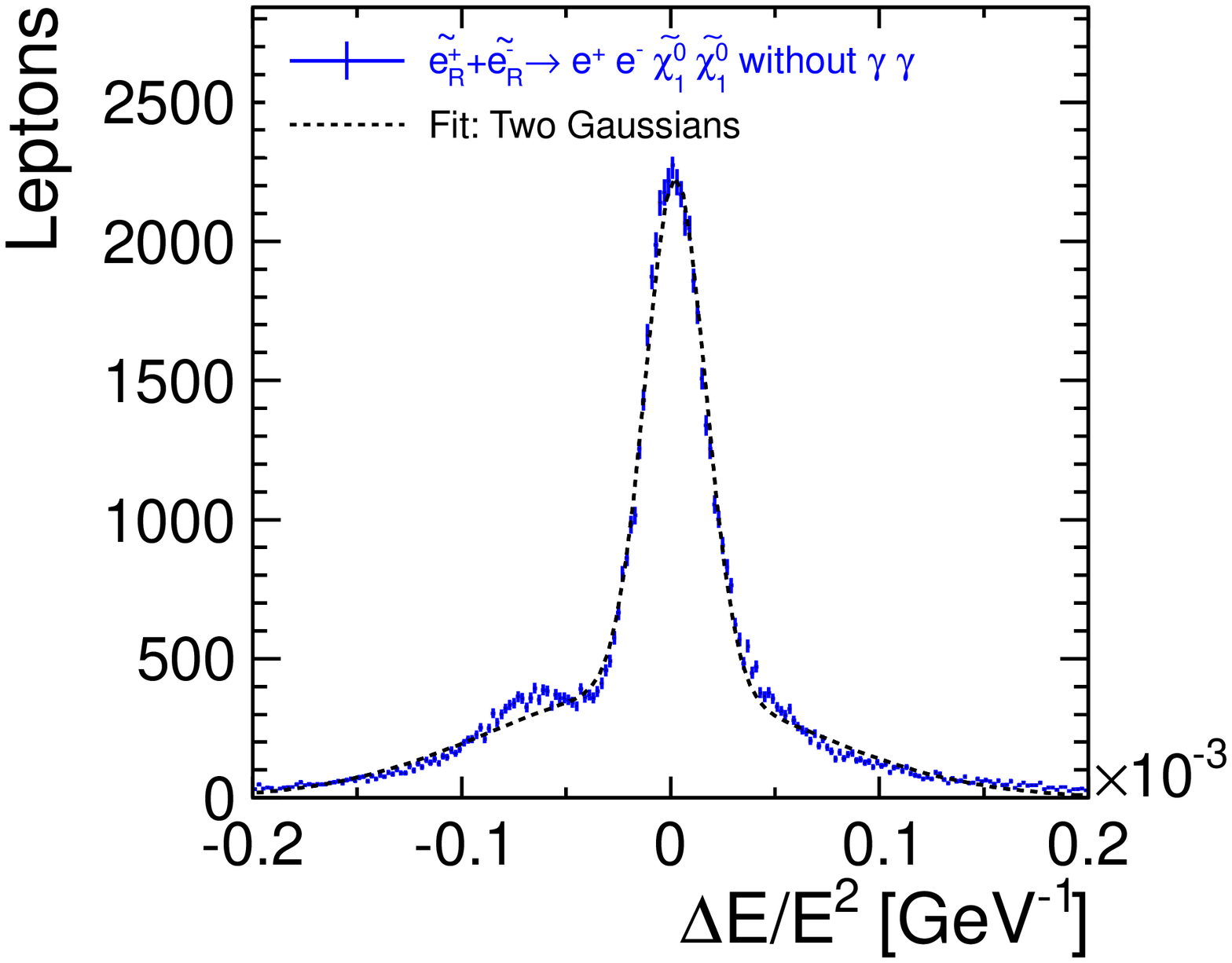}} \\
\hspace{-1.cm}
\subfloat[$\ee \rightarrow  \tilde \nu_e \tilde \nu_e $]
{\includegraphics[width=0.50\textwidth,clip]{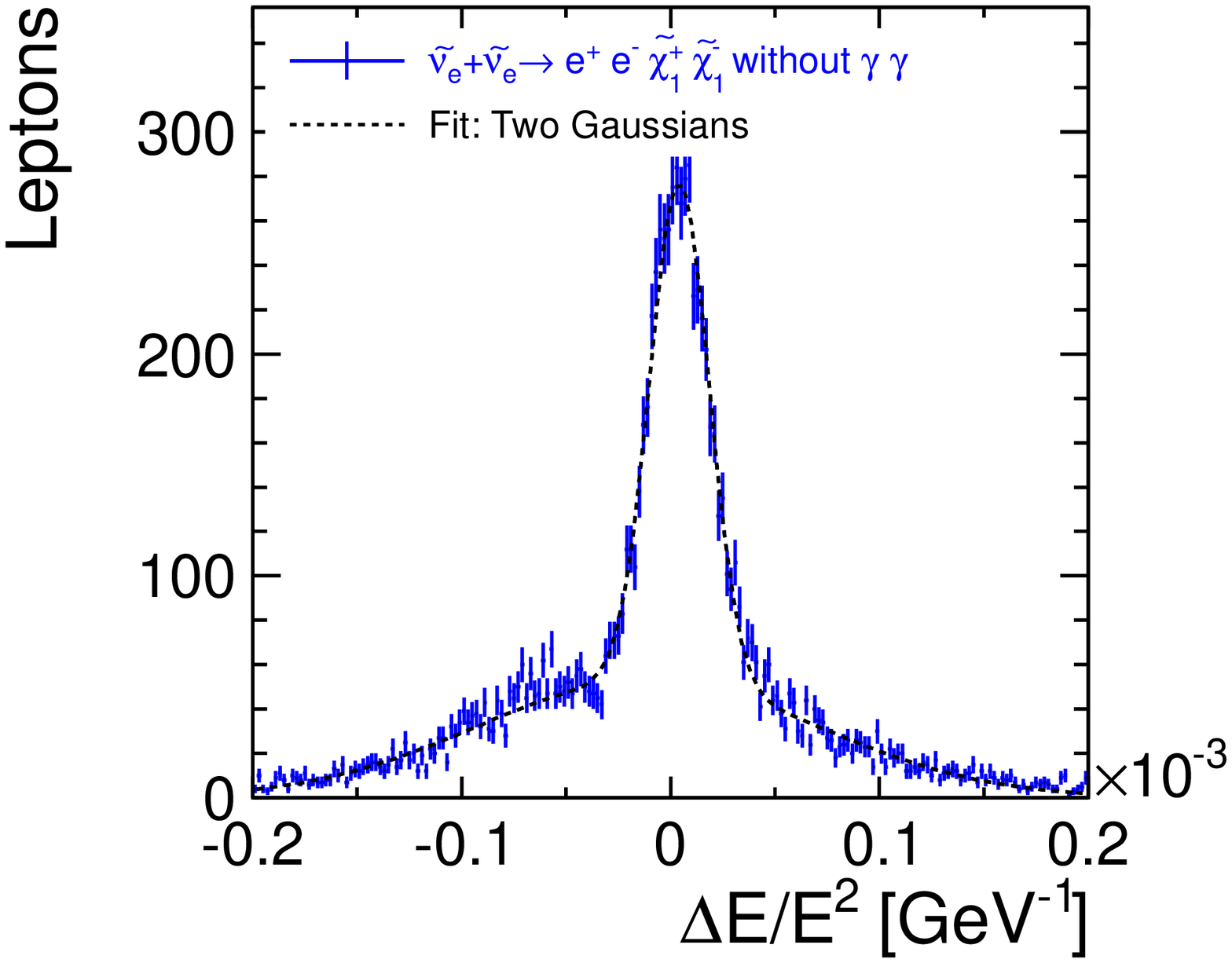}}
\subfloat[$\ee \rightarrow  \tilde e_L^+ \tilde e_L^- $]
{\includegraphics[width=0.50\textwidth,clip]{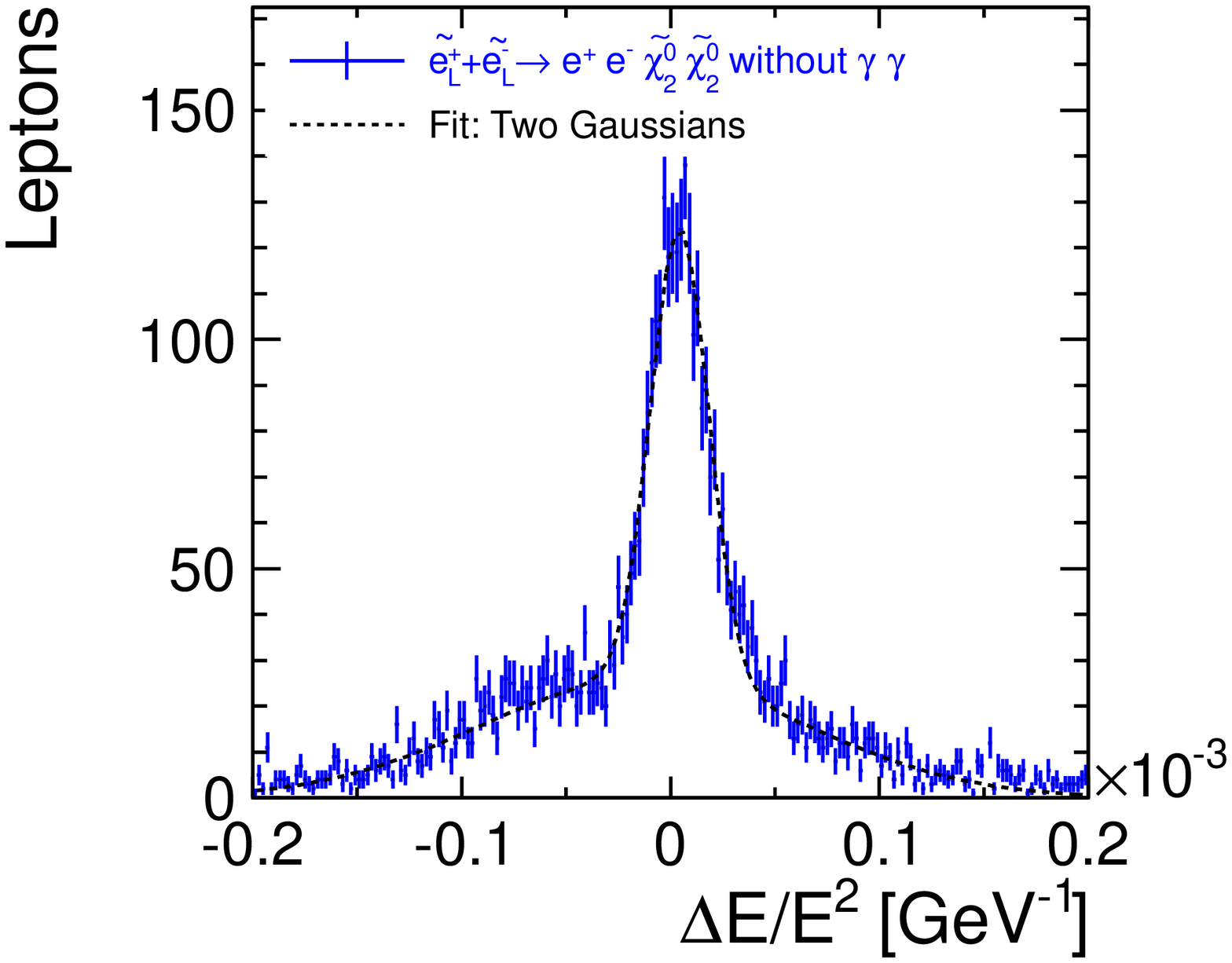}}
\end{tabular}
}
\caption{Lepton energy resolution, at $\sqrt{s}=$ 3 TeV, without $\gamma \gamma \to$ hadron background, for the processes:
$\ee \rightarrow  \tilde \mu_R^+ \tilde \mu_R^- $ (a),
$\ee \rightarrow  \tilde e_R^+ \tilde e_R^- $ (b),
$\ee \rightarrow  \tilde \nu_e \tilde \nu_e $ (c), and
$\ee \rightarrow  \tilde e_L^+ \tilde e_L^- $ (d).  }
\label{fig:H1LEAT2_H1L2A_BX000}
\end{figure}

For the processes with two electrons and four jets, see Figure~\ref{fig:H1LEAT2_H1L2A_BX000} (c) and (d),
despite the presence of four jets, the electron energy resolution 
is consistent with the energy resolution obtained for the isolated electrons process, see
Figure~\ref{fig:H1LEAT2_H1L2A_BX000} (b).
%

%
\begin{table} [tbp!]               
\centering
\caption{ Tight particle flow (PFO) selection cuts for the $\gamma \gamma \to \mathrm{hadron}$ background suppression.
The cut value depends on the particle id, on the detector region and on the $\pT$.  
The same cuts are used at \mbox{$\sqrt{s}=$ 3 TeV} and $\sqrt{s}=$ 1.4 TeV. }
\label{tab:selection_cuts}
\begin{tabular}{ l l l }
\hline
\multicolumn{3}{|c|}{Photons  } \\  \hline
Central region                  & 1.0~GeV~$<\pT~<$4.0~GeV & t~$<$~2.0~nsec \\
$\mathrm {cos\theta~\le~0.975}$ & 0.2~GeV~$\le~\pT~<$1.0~GeV & t~$<$~1.0~nsec \\          
\hline        
Forward regions         & 1.0~GeV~$<~\pT~<$4.0~GeV & t~$<$~2.0~nsec \\
$\mathrm {cos\theta~>~0.975}$   & 0.2~GeV~$\le~\pT~<$1.0~GeV & t~$<$~1.0~nsec \\
\hline
\multicolumn{3}{|c|}{Neutral hadrons  } \\  \hline
Central region                  & 1.0~GeV~$<~\pT~<$8.0~GeV  & t~$<$~2.5~nsec \\
$\mathrm {cos\theta~\le~0.975}$ & 0.5~GeV~$\le~\pT~<$1.0~GeV & t~$<~$1.5~nsec
 \\
\hline
Forward regions         & 1.0~GeV~$<~\pT~<$8.0~GeV  & t~$<$~1.5~nsec \\
$\mathrm {cos\theta~>~0.975}$   & 0.5~GeV~$\le~\pT~<$1.0~GeV & t~$<$~1.0~nsec \\
\hline
\multicolumn{3}{|c|}{Charged particles  } \\  \hline
Central/Forward regions  & 1.0~GeV~$<~\pT~<$4.0~GeV   & t~$<$~2.0~nsec \\
                         & 0.0~GeV~$\le~\pT~<$1.0~GeV & t~$<$~1.0~nsec \\
\hline
\end{tabular}
\end{table}

\subsection{Reconstruction with beam-induced background}
The creation of electron-positron pairs and the production of hadrons in $\gamma\gamma$
interactions are expected to be the dominating source of background events originating from the interaction
region ~\cite{LCD:2011-020}. The beam-beam interaction leading to the production of these background particles
was simulated with the GUINEAPIG program ~\cite{c:thesis}. 
The average number of $\gamma\gamma$ interactions for each bunch crossing
is 3.2 at 3 TeV and 1.3 at \mbox{1.4 TeV}. At 3 TeV the pile-up of this background over the entire 156 ns bunch-train 
deposits 19 TeV of energy in the calorimeters, of which approximately 90\% occurs in the endcap and 10\%
in the barrel regions. On average, there is 1.2 TeV of reconstructed energy from $\gamma\gamma \to$~hadrons that are 
in the same readout window as the physics event. To reduce this energy deposit, $\pT$ and additional timing cuts are applied.
The presence of the $\gamma\gamma \to$~hadron background sets strong requirements
for the design of the CLIC detector and its readout. 

To investigate the effect of beam-induced background, the reconstruction software is run overlaying particles
produced by $\mathrm{\gamma \gamma \rightarrow hadrons}$ interactions \cite{overlay:2011}.
The $\mathrm{\gamma \gamma \rightarrow hadrons}$ event sample was generated with 
{\sc Pythia} and simulated. From this sample we randomly select for each physics event the
equivalent of \mbox{60 bunch} crossings, assuming 3.2 events per bunch crossing at 3 TeV ~\cite{LCD:2011-020} 
and 1.3 events per bunch crossing at \mbox{1.4 TeV}.

The detector hits from these events are merged with those from the physics event before the reconstruction.
A time window of 10 nsec on the detector integration time is applied for all detectors,
except for the HCAL barrel for which the window is 100 nsec.
After particle reconstruction timing cuts in the range of 1 to 3 nsec are applied in order to
reduce the number of particles coming from $\mathrm{\gamma \gamma \rightarrow hadrons}$ interactions
and to optimize the energy resolution.
The cut values vary according to the particle type (photon, neutral hadron, charged particle),
the detector region, (central, forward) and the 
$\pT$ 
of the particle.
Table~\ref{tab:selection_cuts} shows the cut values for the tight particle flow (PFO) selection.

\begin{figure}[ht]
\centering
\resizebox{\textwidth}{!} {
\begin{tabular}{c}
\hspace{-1.cm}
\subfloat[$\ee \rightarrow  \tilde \mu_R^+ \tilde \mu_R^- $, no PFO selection ]
{\includegraphics[width=0.50\textwidth,clip]{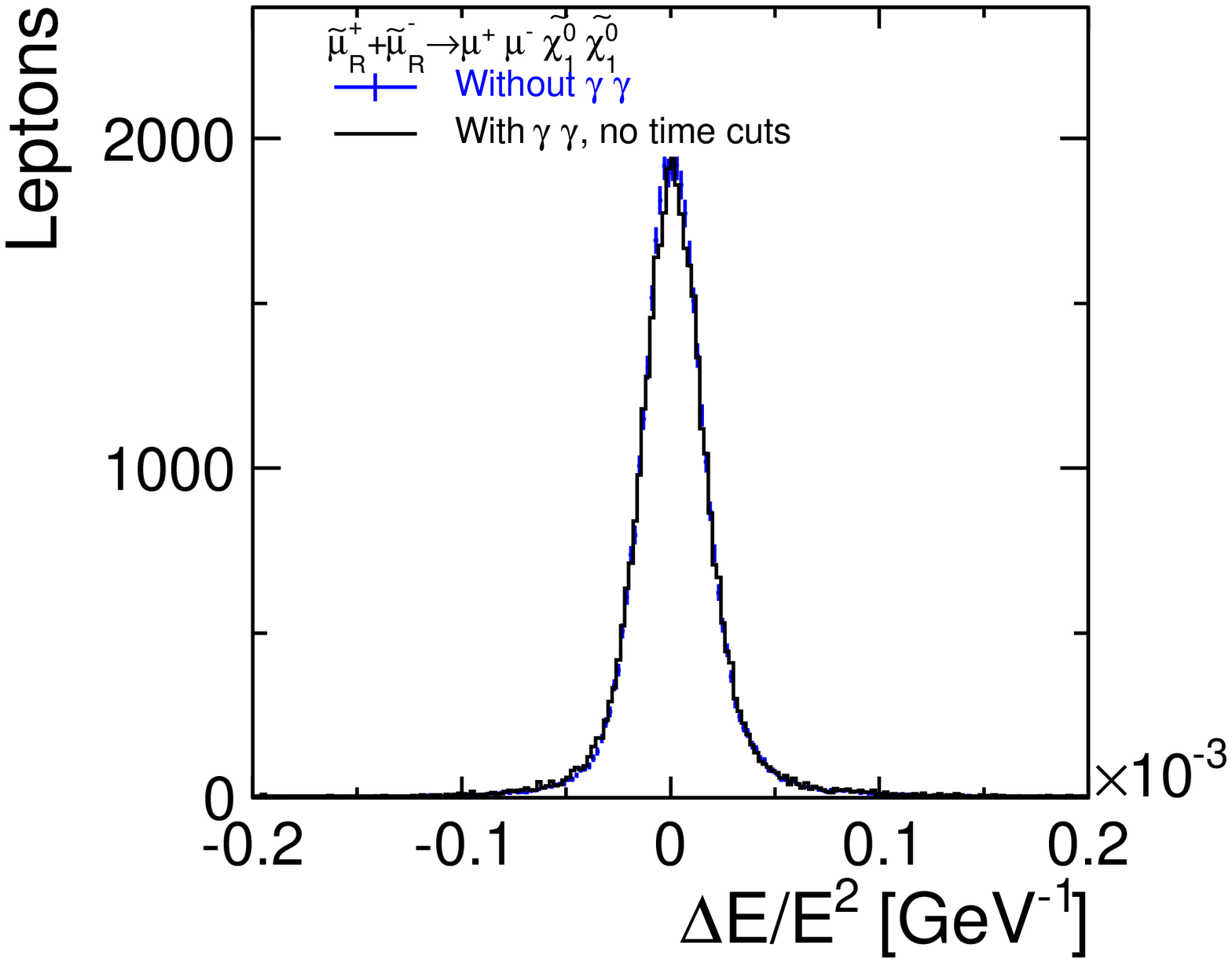}}
\subfloat[$\ee \rightarrow  \tilde e_R^+ \tilde e_R^- $, no PFO selection]
{\includegraphics[width=0.50\textwidth,clip]{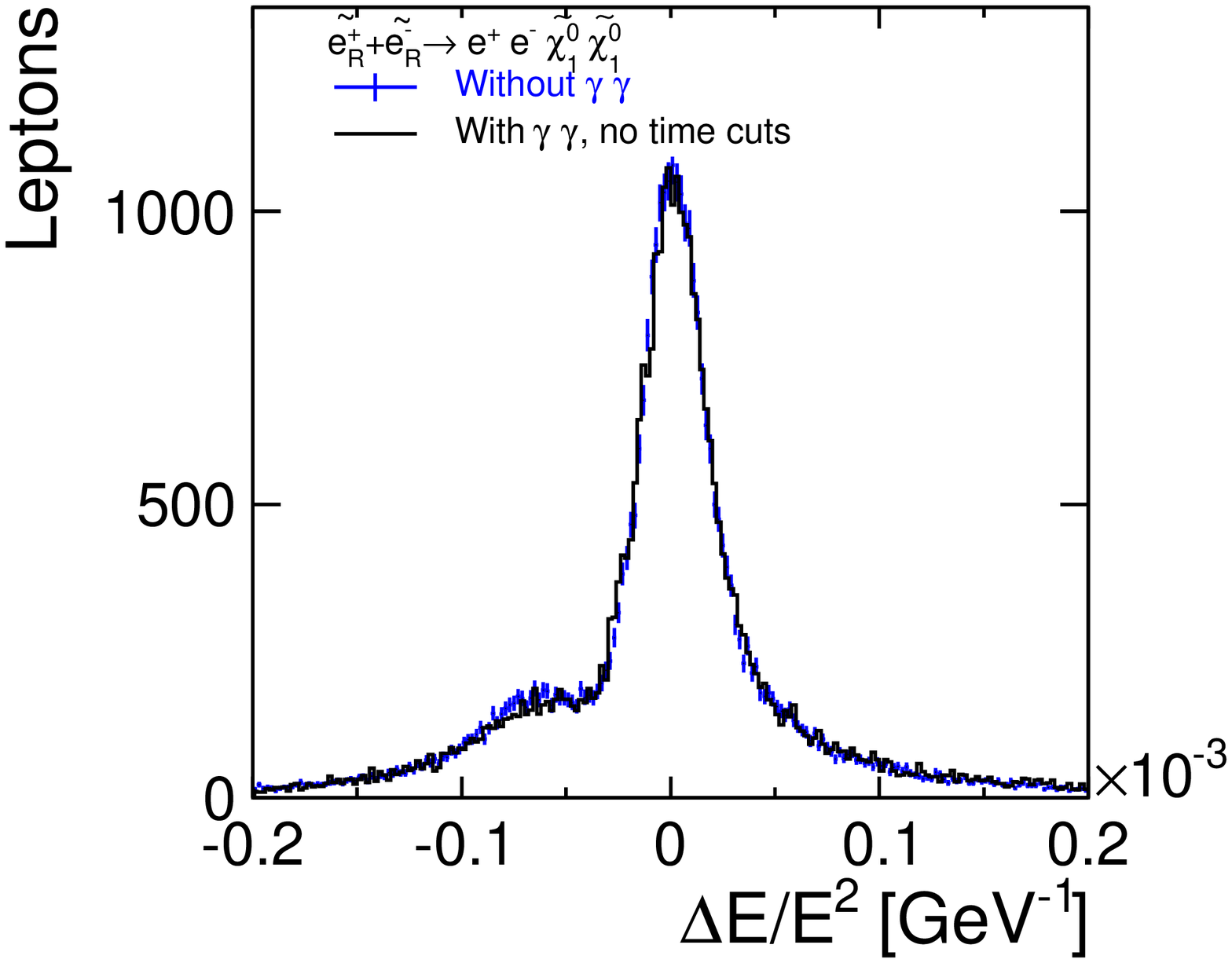}} \\
\hspace{-1.cm}
\subfloat[$\ee \rightarrow  \tilde \nu_e \tilde \nu_e $, no PFO selection]
{\includegraphics[width=0.50\textwidth,clip]{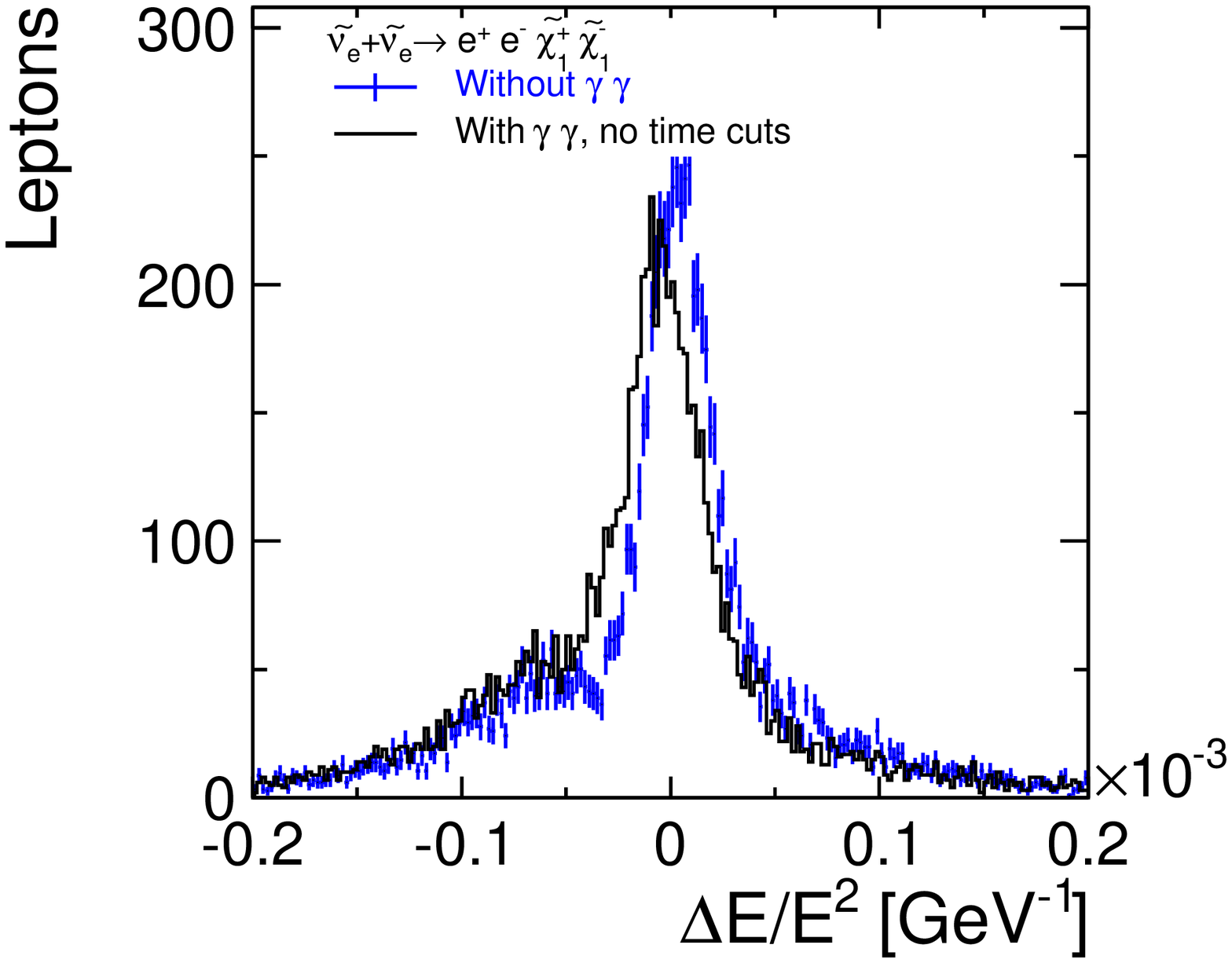}}
\subfloat[ $\ee \rightarrow  \tilde \nu_e \tilde \nu_e $, tight PFO selection]
{\includegraphics[width=0.50\textwidth,clip]{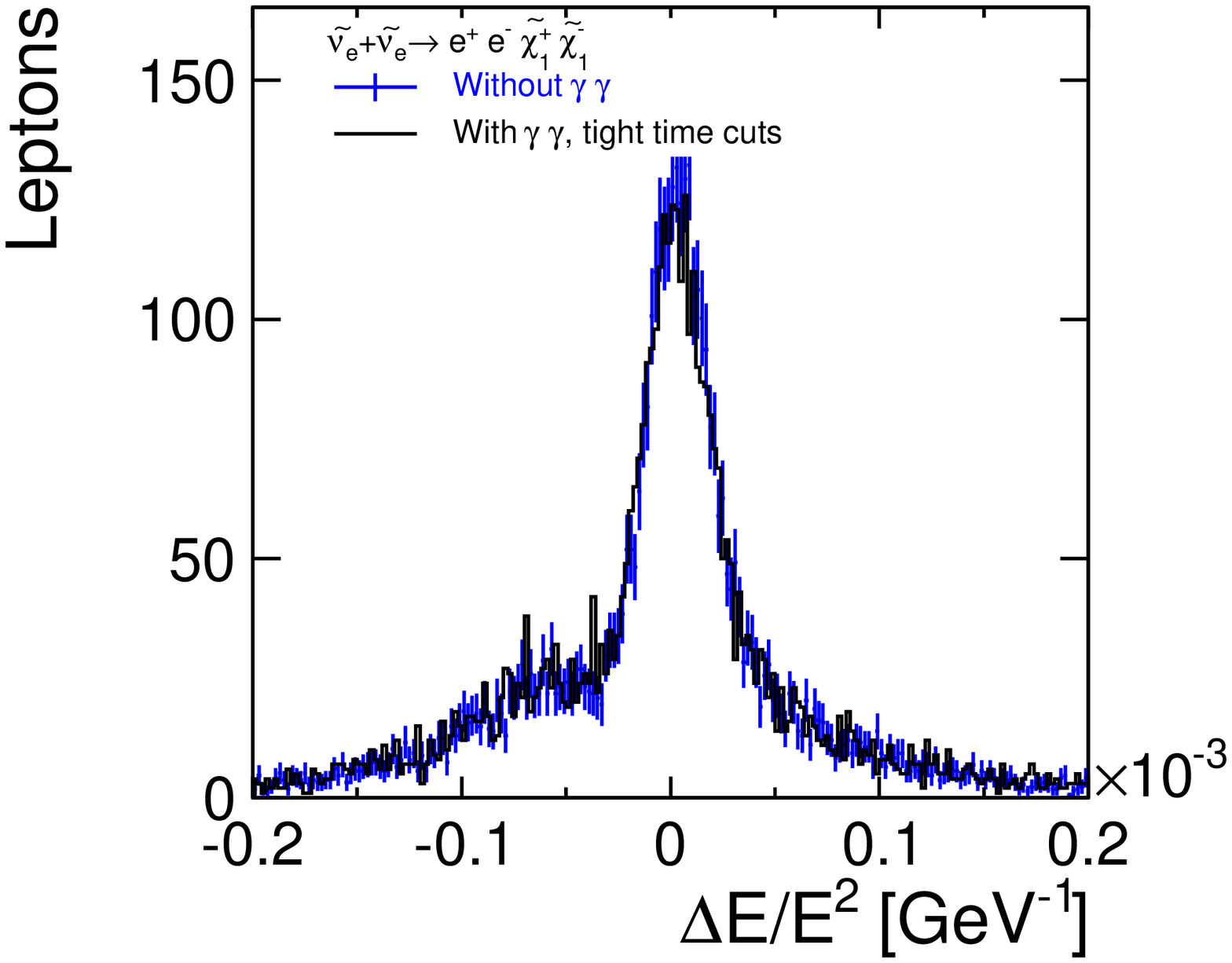}}               
\end{tabular}
}
\caption{Lepton energy resolution, without and with overlaid background at $\sqrt{s}=$ 3 TeV, for the processes:
\mbox{$\ee \rightarrow  \tilde \mu_R^+ \tilde \mu_R^- $, $\pT>4$~GeV; no PFO selection (a)},
\mbox{$\ee \rightarrow  \tilde e_R^+ \tilde e_R^- $, $\pT>4$~GeV; no PFO selection (b)},
\mbox{$\ee \rightarrow  \tilde \nu_e \tilde \nu_e $, no PFO selection (c)}, and
\mbox{$\ee \rightarrow  \tilde \nu_e \tilde \nu_e $, tight PFO selection (d)}.
\label{fig:H1LEAT2_H1L2A_BX060}
}
\end{figure}

Figures~\ref{fig:H1LEAT2_H1L2A_BX060} (a) and (b) show the lepton energy resolution,
without and with $\mathrm{\gamma \gamma \rightarrow hadron}$ background,
for the processes $\ee \rightarrow \tilde  \mu{_R^+} \tilde \mu{_R^-}$ and
$\ee \rightarrow  \tilde e{_R^+} \tilde e{_R^-}$ respectively; 
only a cut requiring $\pT>4$~GeV is applied. 
The lepton energy resolution is preserved; the event selection efficiency is reduced by
1.0\% for
$\ee \rightarrow \tilde  \mu{_R^+} \tilde \mu{_R^-}$ and
3.3\% for
$\ee \rightarrow  \tilde e{_R^+} \tilde e{_R^-}$
, see Table~\ref{tab:RecEffi}.
At \mbox{1.4 TeV} the $\mathrm{\gamma \gamma \rightarrow hadron}$ background is a factor two lower, no selection inefficiency is
induced for the process $\ee \rightarrow \tilde  \mu{_R^+} \tilde \mu{_R^-}$; an  inefficiency of 1.5\% is
induced for the process $\ee \rightarrow  \tilde e{_R^+} \tilde e{_R^-}$.
In final states with four jets and two leptons, the background from  $\gamma \gamma \to $hadrons
cannot be  removed using a similar 
$\pT$ 
cut, as this would 
significantly degrade the jet energy reconstruction.
Figure~\ref{fig:H1LEAT2_H1L2A_BX060} (c) shows the bias in
the reconstructed electron energy when the $\gamma \gamma \to $hadron background is included.
This bias is due to additional background particles being associated with the
electron in the attempt to account for FSR and bremsstrahlung.
Without PFO cuts, the energy resolution is not degraded but the central value is shifted.
Figure~\ref{fig:H1LEAT2_H1L2A_BX060} (d) shows
the lepton energy resolutions without and
with $\gamma \gamma \to $ hadrons overlaid after tight PFO selection cuts. The cuts restore
the central value  and preserve the energy resolution, but reduce the reconstruction efficiency
$\mathrm{ \epsilon_R}$ by 6\%, see Table~\ref{tab:RecEffi}.


\begin{figure}[tbp]
\centering
\resizebox{\textwidth}{!} {
\begin{tabular}{c}
\hspace{-1.cm}
\subfloat[ no PFO selection cuts]
{\includegraphics[width=0.50\textwidth,clip]{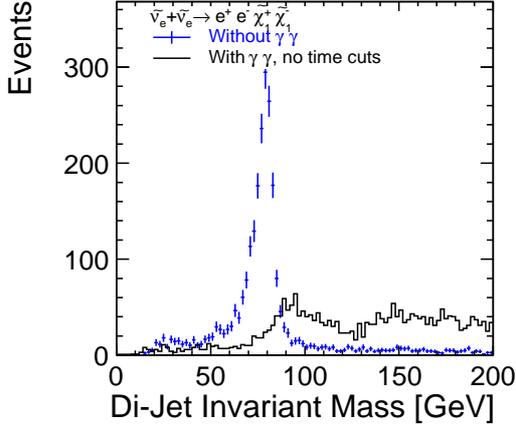}}
\subfloat[ tight PFO selection]
{\includegraphics[width=0.50\textwidth,clip]{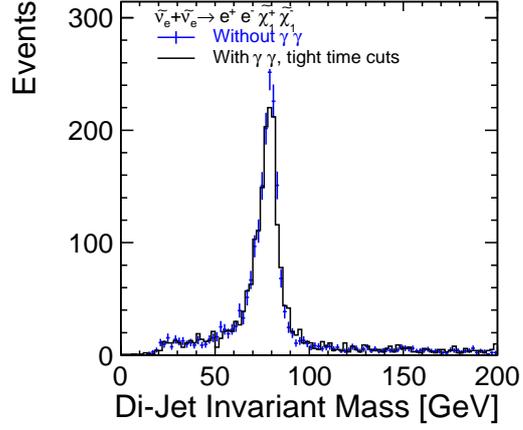}}
\end{tabular}
}
\caption{$W$ boson mass distribution, at $\sqrt{s}=$ 3 TeV,
 for the process $\ee \rightarrow \tilde \nu_e \tilde \nu_e $
 without and with overlaid background:
\ifnote
\else
without PFO cuts (a) and with tight PFO selection (b).
\fi
}
\label{fig:H1RM_BX000_BX060_A}
\end{figure}

\begin{figure}[htbp]
\centering
\resizebox{\textwidth}{!} {
\begin{tabular}{c}
\hspace{-1.cm}
\subfloat[without overlaid background]
{\includegraphics[width=0.50\textwidth,clip]{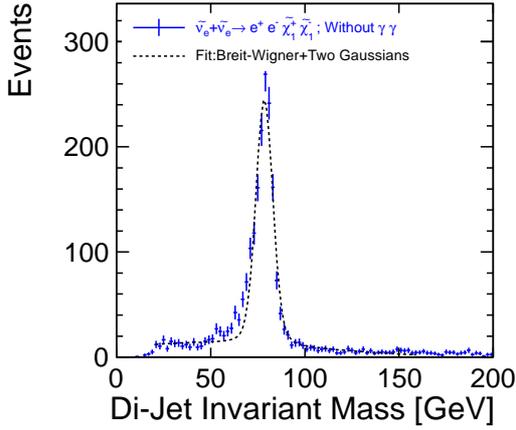}}
\subfloat[with overlaid background and tight selection cuts ]
{\includegraphics[width=0.50\textwidth,clip]{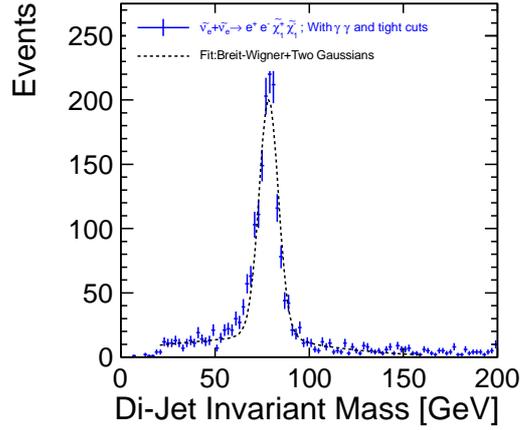}} \\
\end{tabular}
}
\caption{$W$ boson mass distribution fit, at $\sqrt{s}=$ 3 TeV, for the process $\ee \rightarrow \tilde \nu_e \tilde \nu_e $
 without overlaid background (a) and with overlaid background and tight selection cuts (b).}
\label{fig:H1RM_BX000_BX060_B}
\end{figure}

\begin{figure}[htbp]
\centering
\resizebox{\textwidth}{!} {
\begin{tabular}{c}
\hspace{-1.cm}
\subfloat[ $\ee \rightarrow  \tilde \nu_e \tilde \nu_e $ and $\ee \rightarrow  \tilde e_L^+ \tilde e_L^- $ :no b-tag]
{\includegraphics[width=0.50\textwidth,clip]{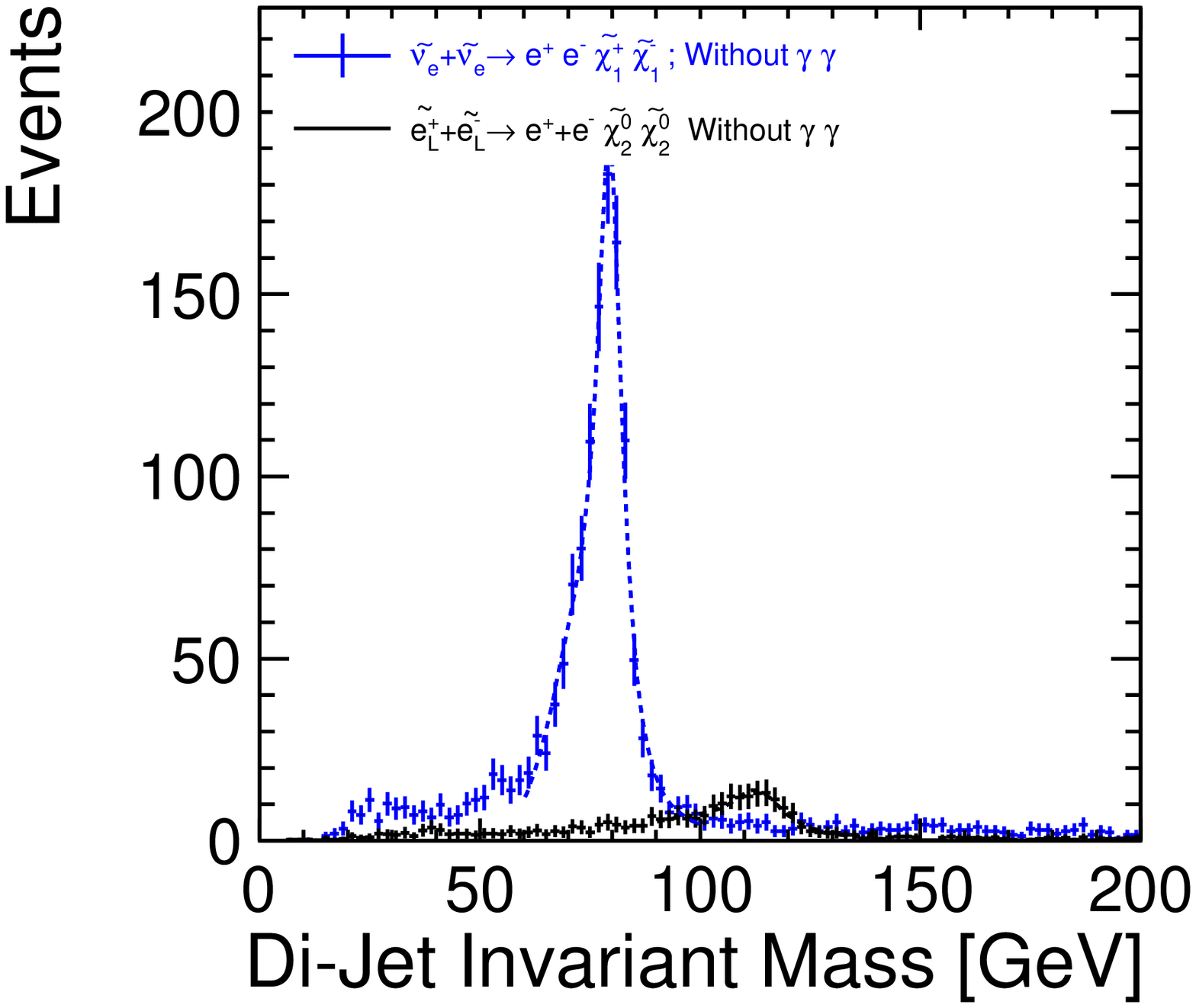}}
\subfloat[ $\ee \rightarrow$~  inclusive SUSY $\rightarrow $ 4 jets:b-tag ]
{\includegraphics[width=0.50\textwidth,height=0.43\textwidth]{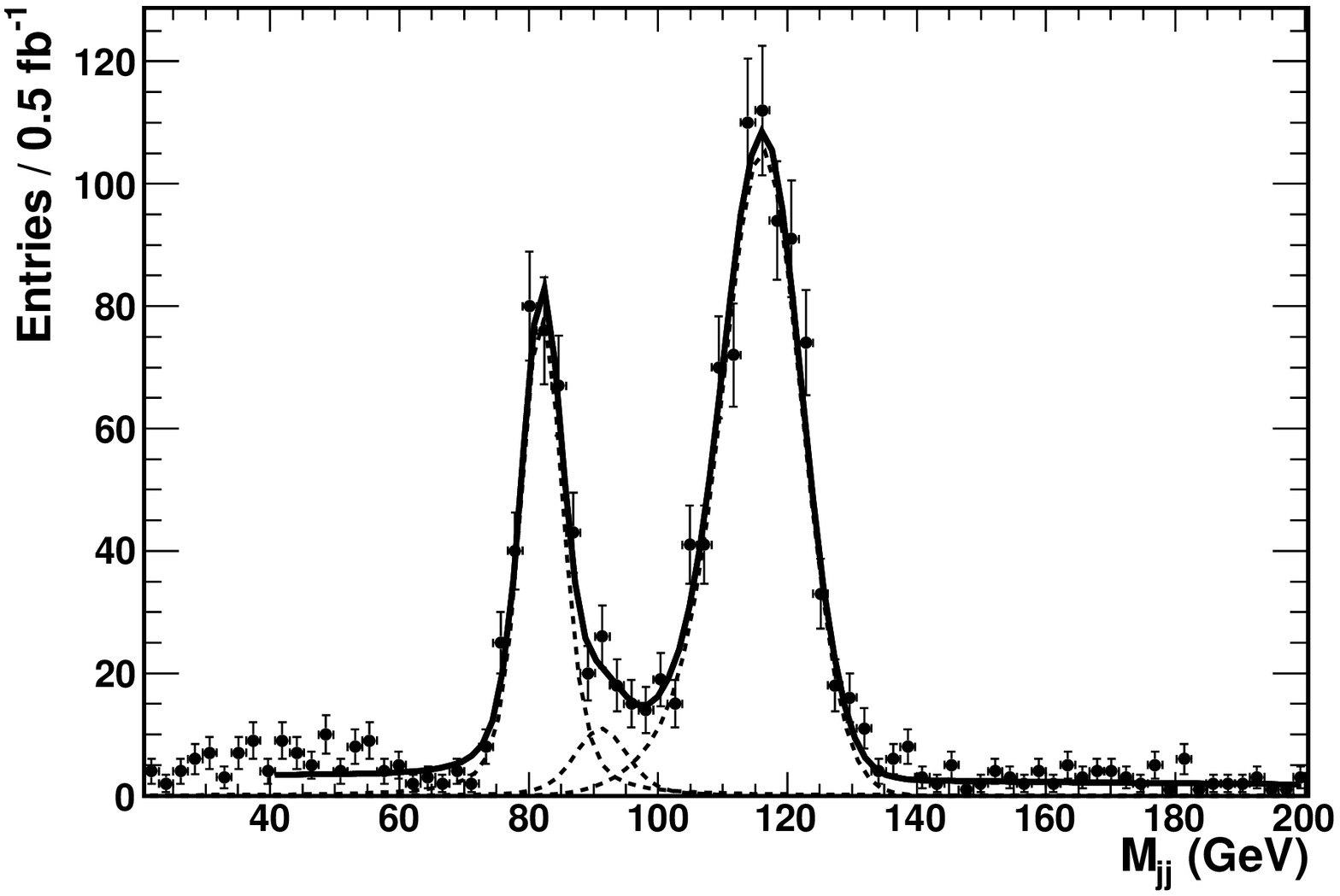}}
\end{tabular}
}
\caption{$W$ and $H$ boson mass distributions, at $\sqrt{s}=$ 3 TeV, for the processes:
$\ee \rightarrow  \tilde \nu_e \tilde \nu_e $ and 
\mbox{$\ee \rightarrow  \tilde e_L^+ \tilde e_L^- $ $\rightarrow $ 2 leptons and 4 jets (a)}, and
\mbox{$\ee \rightarrow$~  inclusive SUSY $\rightarrow $ 4 jets (b).}
}
\label{fig:H1RM_BX000_C}
\end{figure}

Figure~\ref{fig:H1RM_BX000_BX060_A} shows, for the process $\ee \rightarrow  \tilde \nu_e \tilde \nu_e $ at 3 TeV,
the $W$ boson mass distribution without and with overlaid background:
\ifnote
\else
without PFO cuts (a) and with tight PFO selection cuts (b).
\fi
The tight selection cuts give a similar mass distribution as the one obtained without overlaid background.
To estimate the mass resolution degradation,
Figure~\ref{fig:H1RM_BX000_BX060_B} shows the $W$ boson mass distribution fit, for the process $\ee \rightarrow \tilde
\nu_e \tilde \nu_e $ without overlaid background (a) and with overlaid background and tight selection cuts (b).
The mass distributions are fitted with a Breit-Wigner convoluted with two Gaussians, one
Gaussian takes into account the resolution in the peak, the second the tails.
The most probable mass value is fixed as well as the natural width of the $W$.
The width of the peak convoluted Gaussian is 4.1 GeV without overlaid background, it increases to 4.7 GeV with overlaid background
and tight PFO selection cuts.
The fraction of events in the peak gaussian is 90\% without overlaid background and 89\% with overlaid background.
Figure~\ref{fig:H1RM_BX000_C} (a) shows the $W$ and $H$ boson mass distributions for the processes
$\ee \rightarrow \tilde \nu_e \tilde \nu_e $ and $\ee \rightarrow  \tilde e_L^+ \tilde e_L^- $.
The distributions which correspond to an integrated luminosity of 2000 $\mathrm {fb^{-1}}$ are fitted with
two Breit-Wigner functions. The mass distribution
of the Higgs boson is broader than the $W$ one, due to a 10\% background component from $Z$ boson decays
and due to semi-leptonic heavy flavour decays in the $H \to b \bar b$ process.
In this analysis, no flavour tagging is applied.
Figure~\ref{fig:H1RM_BX000_C} (b) shows the boson mass distributions for all inclusive SUSY processes
with four jet final states ~\cite{Alster:2011}. It illustrates that adding $b$-tag information in the analysis
would improve the separation of $W$ and $H$ final states.


\section{Event Selection}
\begin{figure}[tp]
\centering
\resizebox{\textwidth}{!} {
\begin{tabular}{c}
\hspace{-1.cm}
\includegraphics[width=0.85\textwidth]{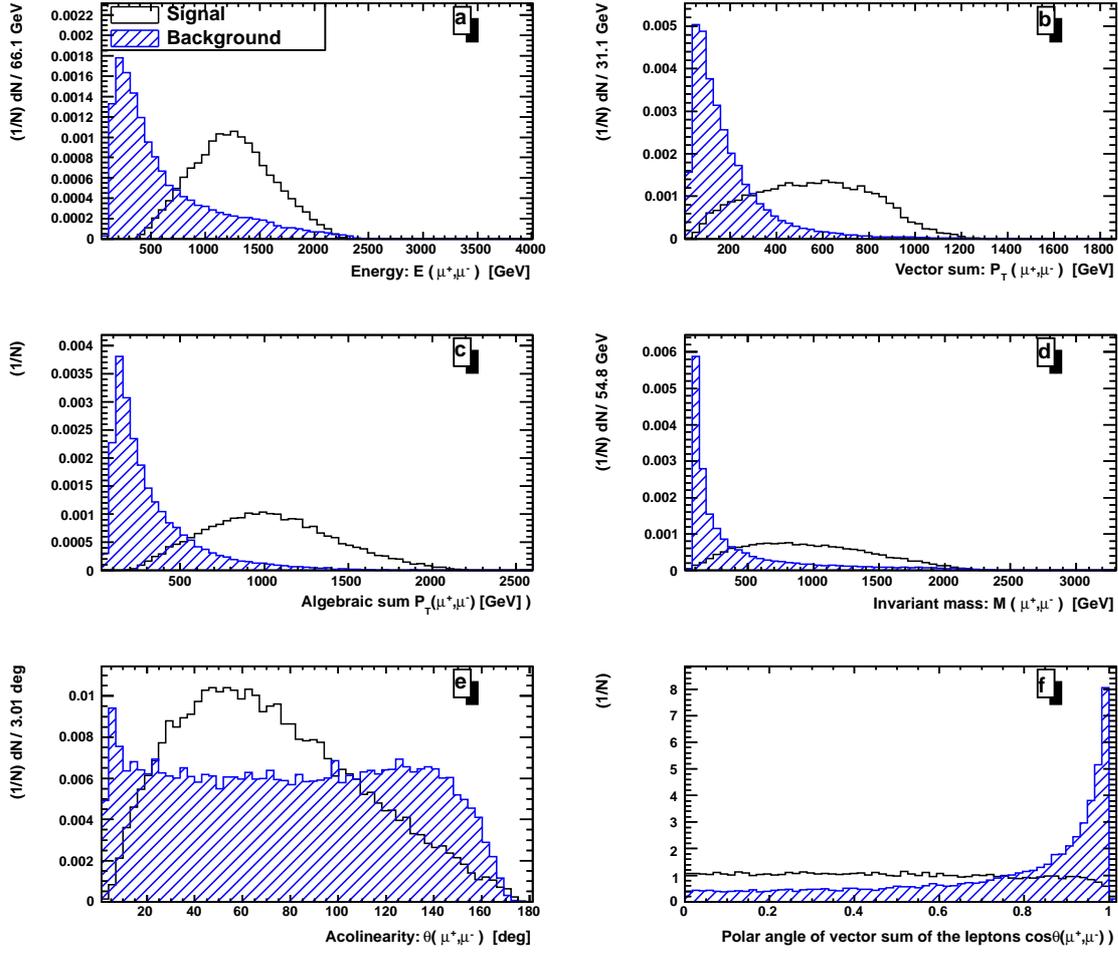}       
\end{tabular}
}   
 \caption{Discriminating variables used to separate signal and background events
for the process \mbox{$\ee \rightarrow  \tilde \mu_R^+ \tilde \mu_R^- $} at $\sqrt{s}=$ \mbox{3 TeV}:
dimuon energy (a),
dimuon $\pT$ vector sum (b),
dimuon $\pT$ algebraic sum (c),
dimuon invariant mass (d),
acolinearity (e) and
polar angle of the vector sum of the leptons (f).
}
\label{fig:205_variables_id_c1}
\end{figure}

All signal processes have two undetected \neutralino{1}'s in the final state.
Therefore, the main characteristics of these events are
missing energy,
missing transverse momentum and acoplanarity.
Despite this signature, the large Standard Model
backgrounds make the analysis rather challenging.
To distinguish signal events from background events the following set of discriminating variables is used:
\begin{itemize}
\item dilepton energy $E(L1)+E(L2)$,
\item vector sum $\pT(L1,L2)$ of the two leptons,
\item algebraic sum $\pT(L1)+\pT(L2)$ of the two leptons,
\item dilepton invariant mass $M(L1,L2)$,
\item dilepton velocity $\beta(L1,L2)$,
\item $\cos \theta(L1,L2)$;  $\theta(L1,L2)$ is the polar angle of the vector sum of the two leptons,
\item dilepton acollinearity $\pi - \theta_2 - \theta_1 $,
\item dilepton acoplanarity $\pi - \phi_2 - \phi_1 $,
\item dilepton energy imbalance $\Delta=|E(L1)-E(L2)|/|E(L1)+E(L2)| $,
\end{itemize}
where $L1$ and $L2$ are the two leptons. 
For illustration,
Figure~\ref{fig:205_variables_id_c1} (a) shows for the process $\ee \rightarrow \tilde \mu{_R^+} \tilde \mu{_R^-}$,
the normalized distributions of some of the observables, for signal and background events, namely
the dimuon energy, the vector sum of the $\pT$ of the leptons, 
the algebraic sum of the $\pT$ of the leptons
the dimuon invariant mass, 
the acolinearity and
the polar angle of the vector sum of the leptons.

\begin{figure}[tp]
\centering
\resizebox{\textwidth}{!} {
\begin{tabular}{c}
\hspace{-1.cm}
\subfloat[With loose BDT cut]{\includegraphics[width=0.50\textwidth]
{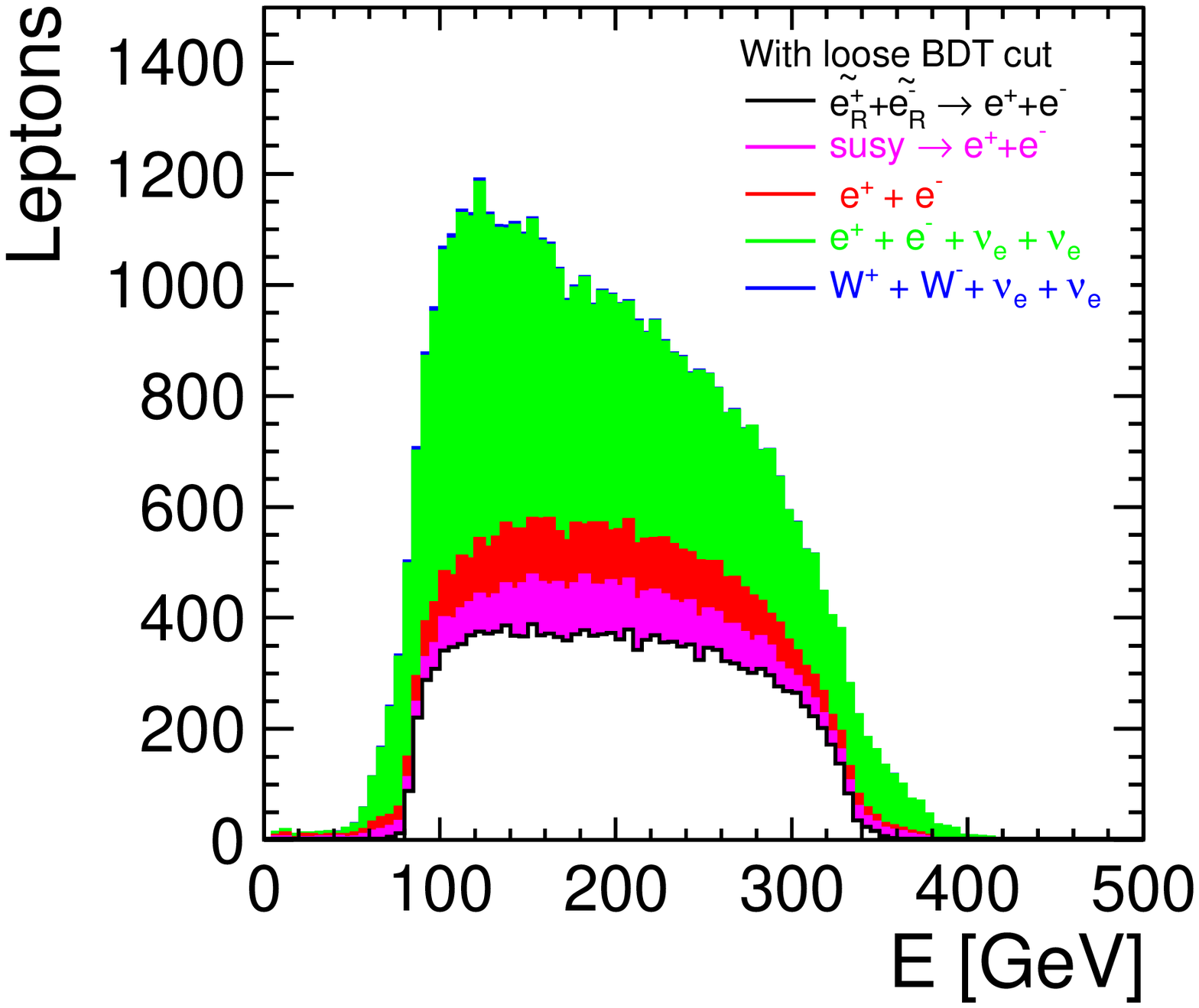}}
\subfloat[With optimized BDT cut]{\includegraphics[width=0.50\textwidth]
{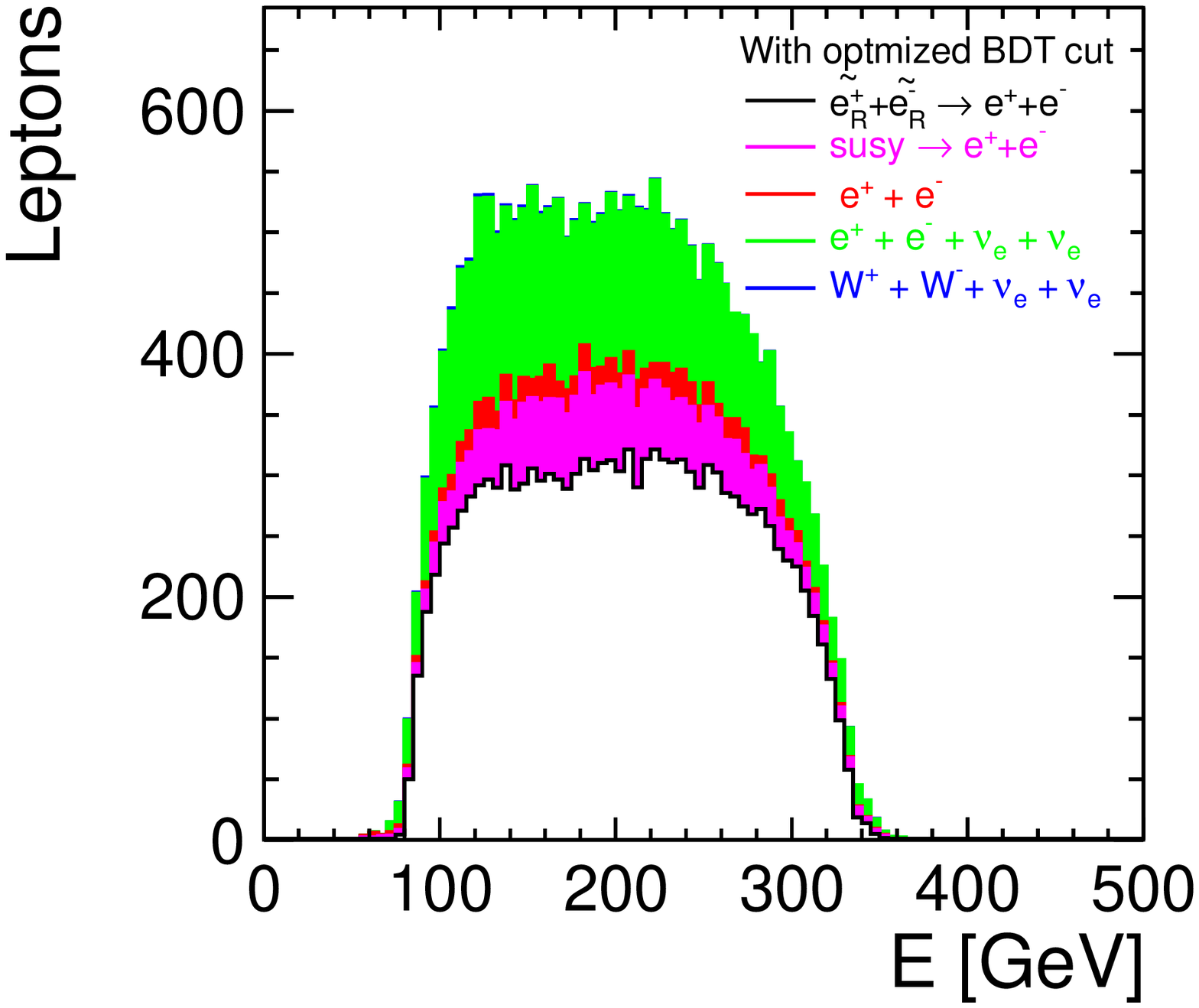}}
\end{tabular}
}  
\caption{Process $\ee \rightarrow  \tilde e_R^+ \tilde e_R^- $ at $\sqrt{s}=$ \mbox{1.4 TeV}:
 electron energy distribution for signal and background events
with loose BDT cut (a), and
with optimized BDT cut (b).
}
\label{fig:202_SPBSTACK}
\end{figure}

The event selection proceeds as follows.
The signal and background samples are split into two equal data size samples
called ``Monte Carlo'' and ``Data''. 
The events of each sample are weighted such that the samples correspond to the same integrated luminosity.
Then the Boosted Decision Trees (BDT) method from the multivariate analysis toolkit, {\sc TMVA}\cite{TMVA:2007},
is used to implement the event selection.
Firstly the discriminating variables of the Monte Carlo sample are input to the BDT method
which trains the BDT probability classifier and computes the weights allowing to distinguish signal from background.
Next the weigths are used to the evaluate the ``Data'' sample, computing
for each event a probability value allowing to rank the events to be signal or background-like.
%
The cut value is chosen to
optimise the significance $ S_{MC} / \sqrt{S_{MC}+B_{MC} } $ versus the signal efficiency
and the background rejection; $S_{MC}$ and $B_{MC}$ are the number of signal and background events
of the MC sample. The cross section and the masses are determined after background subtraction
and efficiency correction; the errors on the masses depend on $\sqrt{ S_{data}+B_{data}+B_{MC} }$.
A stronger BDT cut reduces slightly the significance but decreases significantly the errors
on the masses.  
Figure~\ref{fig:202_SPBSTACK} shows for the process $\ee \rightarrow  \tilde e_R^+ \tilde e_R^- $ at \mbox{$\sqrt{s}=$ 1.4 
TeV}, the stacked electron energy distribution for signal and background events
with a loose BDT cut (a), and with an optimized  BDT cut (b).
At 3 TeV the BDT selection efficiency is 95\% for the dimuon events, 90\%
for the dielectron events and 94\% for the dielectron and four jet events.
At \mbox{1.4 TeV}
the efficiency is 90\% for the dimuon events, 80\% for the dielectron events, 
and 90\% for the dielectron and four jet events.
%
%

\section{Slepton and Gaugino Mass Determination}
After the final selection, the slepton, neutralino or chargino masses are extracted from the position
of the kinematic  edges of the
lepton energy distribution, a technique first proposed for squarks~\cite{Feng:1993sd}, then
extensively applied to sleptons~\cite{Martyn:1999tc}:
\begin{eqnarray}
m_{\tilde \ell^{\pm}}=\frac{\sqrt{s}}{2} \left(1-\frac{( E_{H}-E_{L} )^{2}}{( E_{H}+E_{L})^{2}} \right)^{1/2}
\hspace{0.2cm} \mathrm {and} \hspace{0.2cm}
m_{\neutralino{1}}~\mathrm{or}~m_{\chargino{\pm}}=m_{\tilde \ell^{\pm}} \left( 1-\frac{ 2 (E_{H}+E_{L})}{\sqrt{s}}
\right)^{1/2},
\label{formula:m1m2}
\end{eqnarray}

\noindent where $E_{L}$ and $E_{H}$ are the low and high edges of the lepton energy distribution

\begin{eqnarray}
E_{H,\;L}=\frac{\sqrt{s}}{4}\left( 1- \frac { m_{\neutralino{1}}^{2} } { m_{\tilde \ell{\pm}}^{2} }  \right)
\left( 1 \pm \sqrt{1 - 4  \frac {m_{\tilde \ell{\pm}}^{2}} {s}} \right).
\label{formula:eleh}
\end{eqnarray}

The masses are determined using a three-parameter fit to the
background subtracted energy distribution, 
with $\sigma_{\tilde\ell^{\pm}},~ m_{\tilde\ell^{\pm}} $ and $m_{\neutralino{1}}$ or  $m_{\chargino{\pm}}$  as parameters.
The background subtraction is done using the "Monte Carlo" event sample used to train
the classifier.    
The fit is performed with the {\sc Minuit} minimization  package~\cite{James:1975dr}.
The fit function is:
\begin{eqnarray}
f(E) = \int^{\sqrt{s_{max}}}_{\sqrt{s_{min}}} L_{Eff}(\sqrt{s}) \cdot \int^{E_H(\sqrt{s})}_{E_L(\sqrt{s})}   
U(\sigma_{\tilde\ell^{\pm}}, m_{\tilde \ell^{\pm}}, m_{\neutralino{1}}, \sqrt{s}, E-\tau) \cdot D(\tau) ~d\sqrt{s}~d\tau        
\label{formula:fitfunc}	
\end{eqnarray}

\begin{figure}[htbp]
\centering
\resizebox{\textwidth}{!} {
\begin{tabular}{c}
\hspace{-1.cm}
\subfloat[ $\ee \rightarrow  \tilde e_R^+ \tilde e_R^- $~$\sqrt{s}=$ 3 TeV]
{\includegraphics[width=0.50\textwidth,clip]
{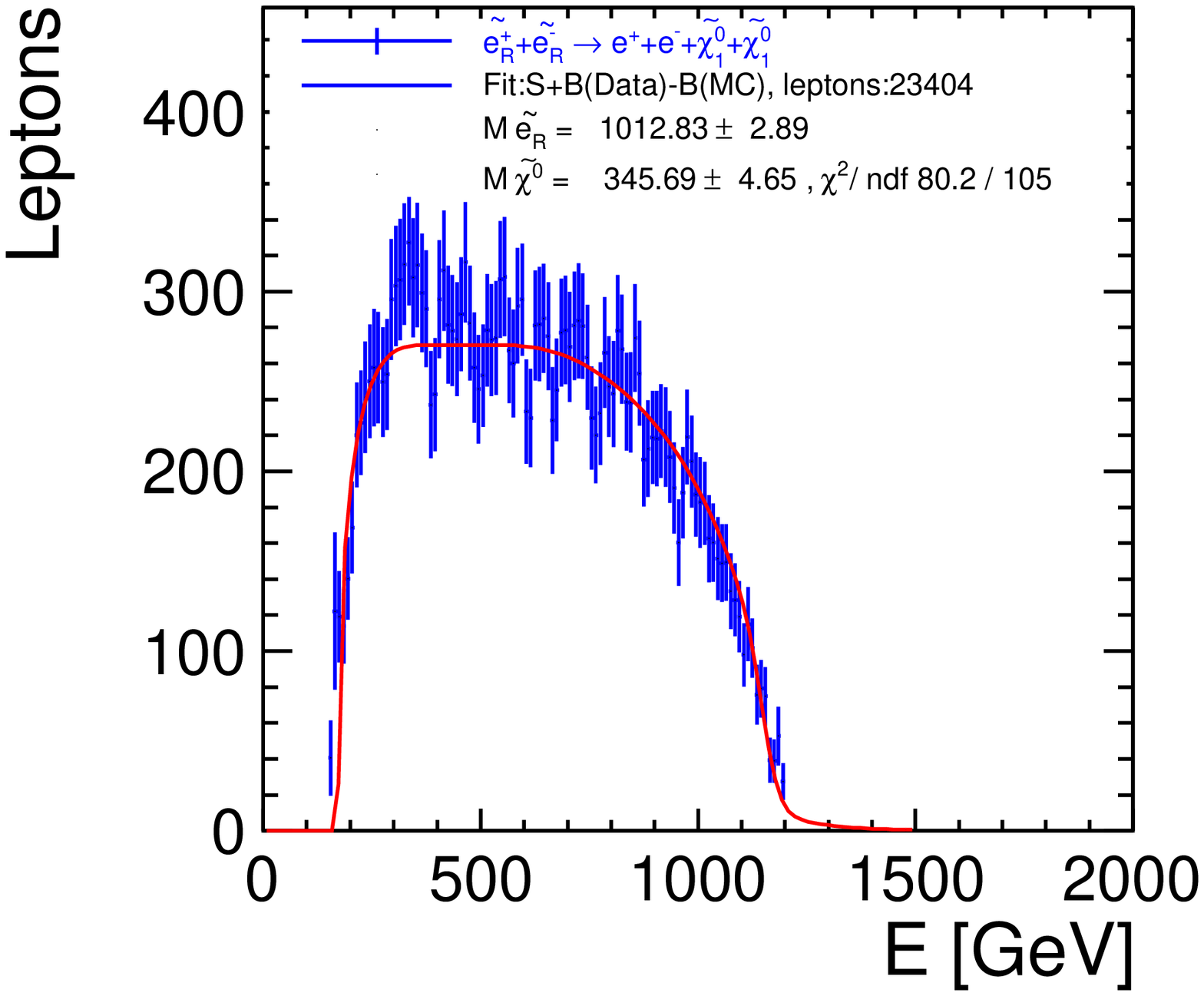}} 
\subfloat[$\ee \rightarrow  \tilde \nu_e \tilde \nu_e $~$\sqrt{s}=$ 3 TeV]
{\includegraphics[width=0.50\textwidth,clip]
{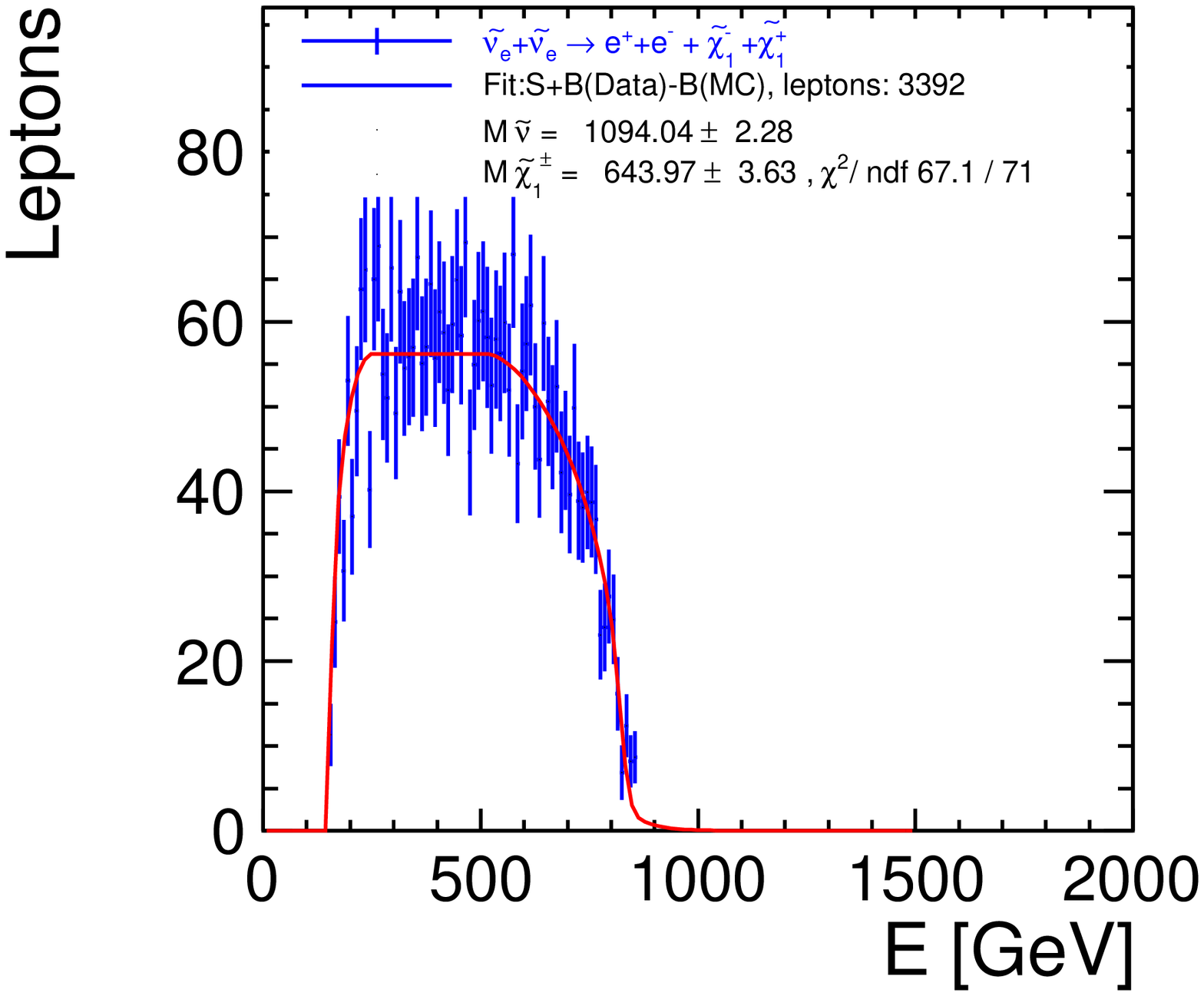}}
\end{tabular}
}
\caption{Lepton energy spectrum and fit results, for the processes:
$\ee \rightarrow  \tilde e_R^+ \tilde e_R^- $ (a) and
$\ee \rightarrow  \tilde \nu_e \tilde \nu_e $ (c) at $\sqrt{s}=$ 3 TeV.
}
\label{fig:H1LEA_FIT}
\end{figure}

\noindent $L_{Eff}(\sqrt{s})$ is the effective luminosity function,
\mbox{$L_{Eff}(\sqrt{s})= L(\sqrt{s}) \otimes ISR(\sqrt{s}) \otimes \sigma_{\tilde\ell^{\pm}}(\sqrt{s})$}.
$L(\sqrt{s})$ is the luminosity spectrum prior to initial state radiation (ISR),
$ISR(\sqrt{s})$ is the $\sqrt{s}$ variation due to ISR
and $\sigma_{\tilde\ell^{\pm}}(\sqrt{s})$ is the slepton cross section.
\textit{U} is a uniform distribution of E, and depends on the process cross section $\sigma_{\tilde\ell^{\pm}}$, 
the slepton and gaugino masses and $\sqrt{s}$ ; the boundaries $E_L, E_H$ of \textit{U} are given by \ref{formula:eleh}.
\textit{D} is the detector resolution function
obtained from the fits shown in
Figure~\ref{fig:H1LEAT2_H1L2A_BX000}.
%
%
Figure~\ref{fig:H1LEA_FIT}
shows, for the processes 
$\ee \rightarrow \tilde e{_R^+} \tilde e{_R^-}$ (a) and
$\ee \rightarrow  \tilde \nu_e \tilde \nu_e $ (b) at \mbox{$\sqrt{s}=$ 3 TeV} 
the lepton energy distributions and fit results.
Table~\ref{tab:results3tev} shows the values of the measured slepton cross sections, slepton masses,
and gaugino masses at  $\sqrt{s}=$ 3 TeV, assuming 2~ab$^{-1}$ of integrated luminosity.
For the process
$e^+e^- \to \tilde e_L^+ ~\tilde e_L^- \to e^+~e^- ~\neutralino{2}~\neutralino{2} $,
the cross section is determined from the fit to the boson mass distribution,
Figure~\ref{fig:H1RM_BX000_C}.
Table~\ref{tab:results1.4tev} shows the results at 1.4 TeV, assuming 1.5~ab$^{-1}$ of integrated luminosity.

\begin{table} [htbp]
\centering
\caption{Values of cross sections, slepton and gaugino masses, and statistical accuracies assuming an integrated
luminosity of 2 $\mathrm {ab^{-1}}$ at  $\sqrt{s}=$ 3 TeV. }
\label{tab:results3tev}
\begin{tabular}{ l l c c c }
\hline
Process &~~Decay Mode &~~ $\sigma $ &~~$m_{\tilde \ell}$  &~~ $m_{\neutralino{1}}$ or $m_{\chargino{\pm}}$  \\
        &             &~~ fb         &~~GeV         & GeV            \\ \hline
$\ee \rightarrow  \tilde \mu_R^+ \tilde \mu_R^- $ &~~ $ \mu^+  \mu^- \neutralino{1} \neutralino{1}$
&~~0.73 $\pm$ 0.02 &~~1011.9 $\pm$ 4.9 &~~342.7 $\pm$ 9.7 \\
$\ee \rightarrow  \tilde e_R^+ \tilde e_R^- $ &~~ $ e^+  e^- \neutralino{1} \neutralino{1}$
&~~6.23 $\pm$ 0.04 &~~1012.8 $\pm$ 2.9 &~~345.7 $\pm$ 4.6 \\
$\ee \rightarrow  \tilde e_L^+ \tilde e_L^- $ &~~ $ e^+  e^- \neutralino{2} \neutralino{2} $
&~~2.77 $\pm$ 0.20 &~~ &~~ \\
$\ee \rightarrow  \tilde \nu_e \tilde \nu_e $
&~~ $ e^+  e^- \chargino{\pm} \chargino{\pm} $
&~13.27 $\pm$ 0.23 &~~1094.0 $\pm$ 2.3 &~~644.0 $\pm$ 3.6 \\
\hline
\end{tabular}
\end{table}
\begin{table} [htbp]
\caption{Values of cross sections, slepton and gaugino masses, and statistical accuracies assuming an integrated
luminosity of 1.5 $\mathrm {ab^{-1}}$ at  $\sqrt{s}=$ 1.4 TeV.}
\label{tab:results1.4tev}
\centering
\begin{tabular}{ l l c c c }
\hline
Process &~~Decay Mode &~~ $\sigma $ &~~$m_{\tilde \ell}$  &~~ $m_{\neutralino{1}}$ or $m_{\chargino{\pm}}$  \\
        &             &~~ fb         &~~GeV         & GeV            \\ \hline
$\ee \rightarrow  \tilde \mu_R^+ \tilde \mu_R^- $ &~~ $ \mu^+  \mu^- \neutralino{1} \neutralino{1}$
&~~1.51 $\pm$ 0.03 &~~559.1 $\pm$ 0.4  &~~357.1 $\pm$ 0.7 \\
$\ee \rightarrow  \tilde e_R^+ \tilde e_R^- $ &~~ $ e^+  e^- \neutralino{1} \neutralino{1}$
&~~5.99 $\pm$ 0.05 &~~557.9 $\pm$ 0.6 &~~356.1 $\pm$ 0.9 \\
$\ee \rightarrow  \tilde \nu_e \tilde \nu_e $
&~~ $ e^+  e^- \chargino{\pm} \chargino{\pm} $
&~~5.13 $\pm$ 0.19 &~~644.5 $\pm$ 2.2 &~~488.8 $\pm$ 1.1 \\
\hline
\end{tabular}
\end{table}

%
\section{Systematic Uncertainty related to the event selection}
For the event selection described in section 4 the signal sample, used to train the classifier
allowing to distinguish signal events from background events,  
was generated with the same slepton and gaugino masses as the data sample. 
With real data the masses are unknown.
In this section we describe the procedure allowing to
determine the masses and assess the error on the masses introduced when the MC masses
are different from the true masses; the evaluation is done for the
process $\ee \rightarrow  \tilde \mu_R^+ \tilde \mu_R^- $ at 1.4 TeV.

\begin{figure}[tp]
\centering
\resizebox{\textwidth}{!} {
\begin{tabular}{c}
\hspace{-1.cm}
\subfloat[BDT trained wither lower signal masses]
{\includegraphics[width=0.50\textwidth]{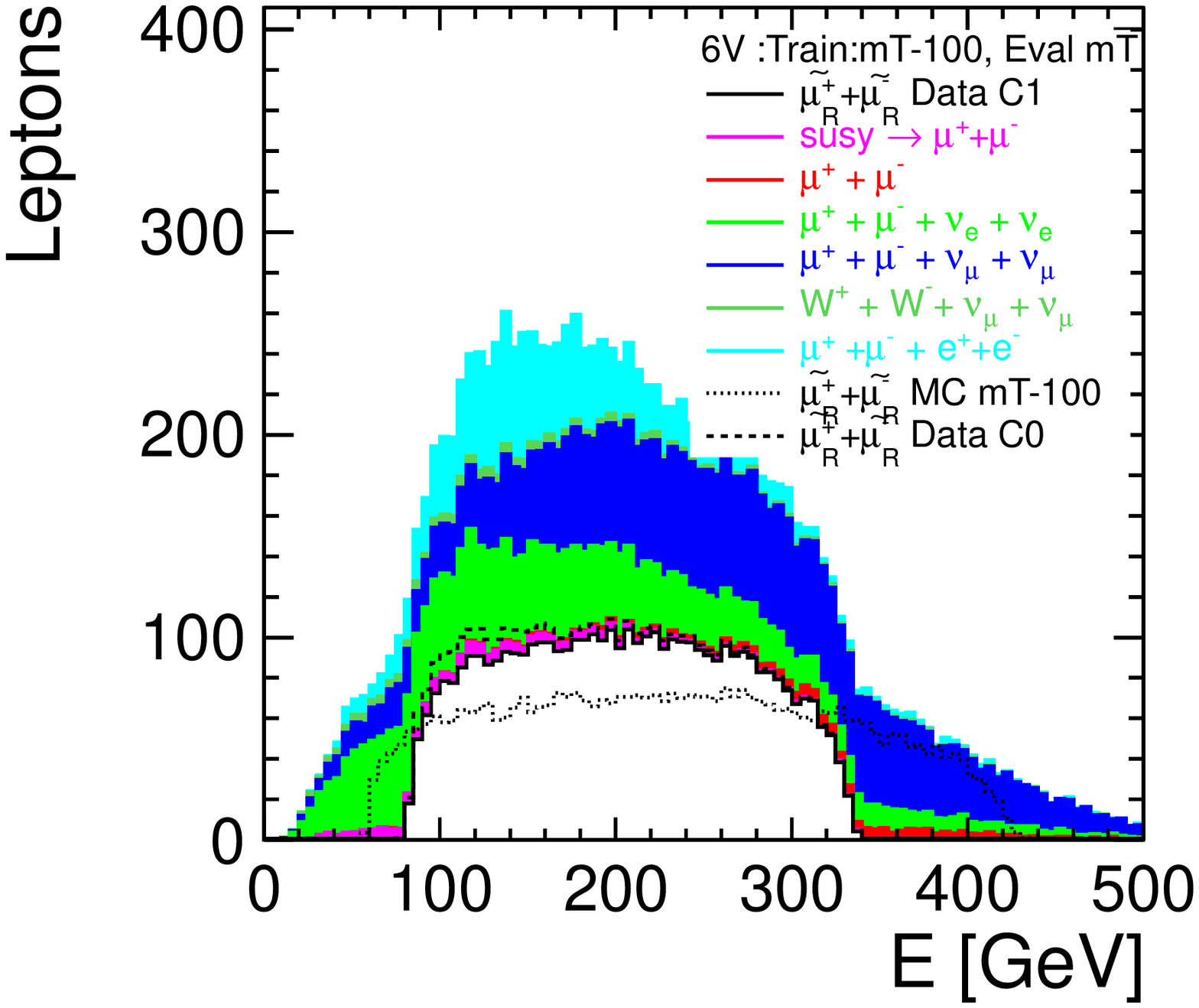}}
\subfloat[BDT trained wither larger signal masses]
{\includegraphics[width=0.50\textwidth]{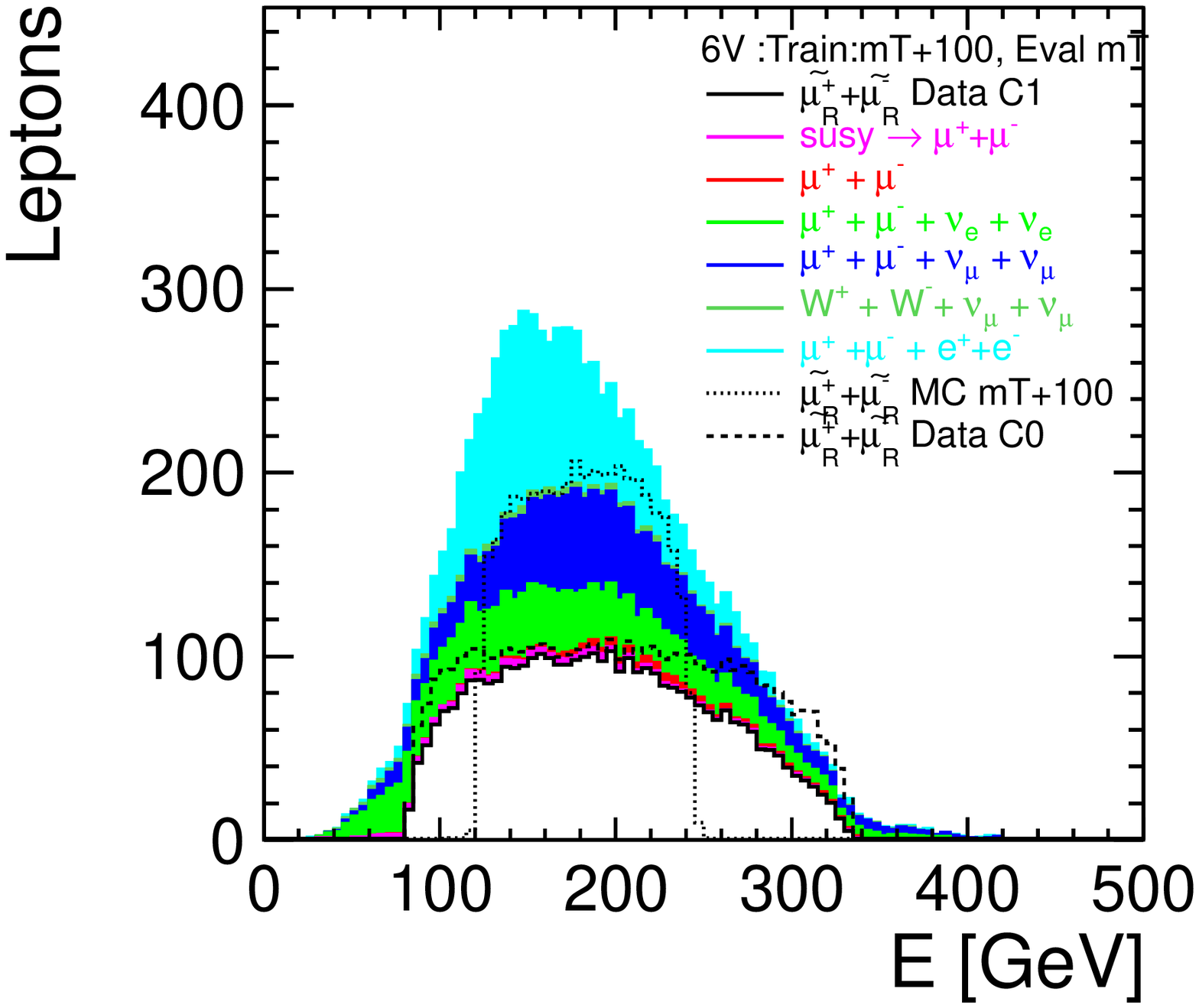}}
\end{tabular}
}  
 \caption{Process $\ee \rightarrow  \tilde \mu_R^+ \tilde \mu_R^- $ at $\sqrt{s}=$ \mbox{1.4 TeV}:
stacked muon energy distribution for signal and background events
with BDT trained wither lower signal masses (a), and
with BDT trained wither larger signal masses (b).
}
\label{fig:205_mT-100_mT+100}
\end{figure}

Firstly signal events for lower smuon and neutralino masses are simulated and reconstructed; 
the smuon and neutralino masses are 459 GeV and 257 GeV respectively, that is to say, data masses - 100 GeV. 
These events are used to train a classifier in which three variables are removed, namely the
dilepton energy, dilepton velocity and dilepton energy imbalance. These variables are most correlated
with the masses. The 6 variables classifier is then used to select the events.
Figure~\ref{fig:205_mT-100_mT+100} (a) shows the the stacked muon energy distribution for signal and background events
selected with the 6 variables classifier trained with masses lower by 100 GeV.
The energy distribution of the MC training sample is the black dotted line;
the energy distribution of the selected signal data sample is the black full line;
and the energy distribution of the signal data sample without selection is the black dashed line.

Next signal events for larger smuon and neutralino masses are simulated and reconstructed. 
The smuon and neutralino masses are 659 GeV and 457 GeV respectively, this is to say, data masses + 100 GeV . 
These events are used to train the classifier which is then used to select the events.
Figure~\ref{fig:205_mT-100_mT+100} b) shows the the stacked muon energy distribution for signal and background events
selected with a classifier trained masses with masses larger by 100 GeV.

\begin{figure}[htbp]
\centering
\resizebox{\textwidth}{!} {
\begin{tabular}{c}
\hspace{-1.cm}
\subfloat[Events selected with a classifier trained with signal masses lower by 10 GeV.]
{\includegraphics[width=0.50\textwidth]
{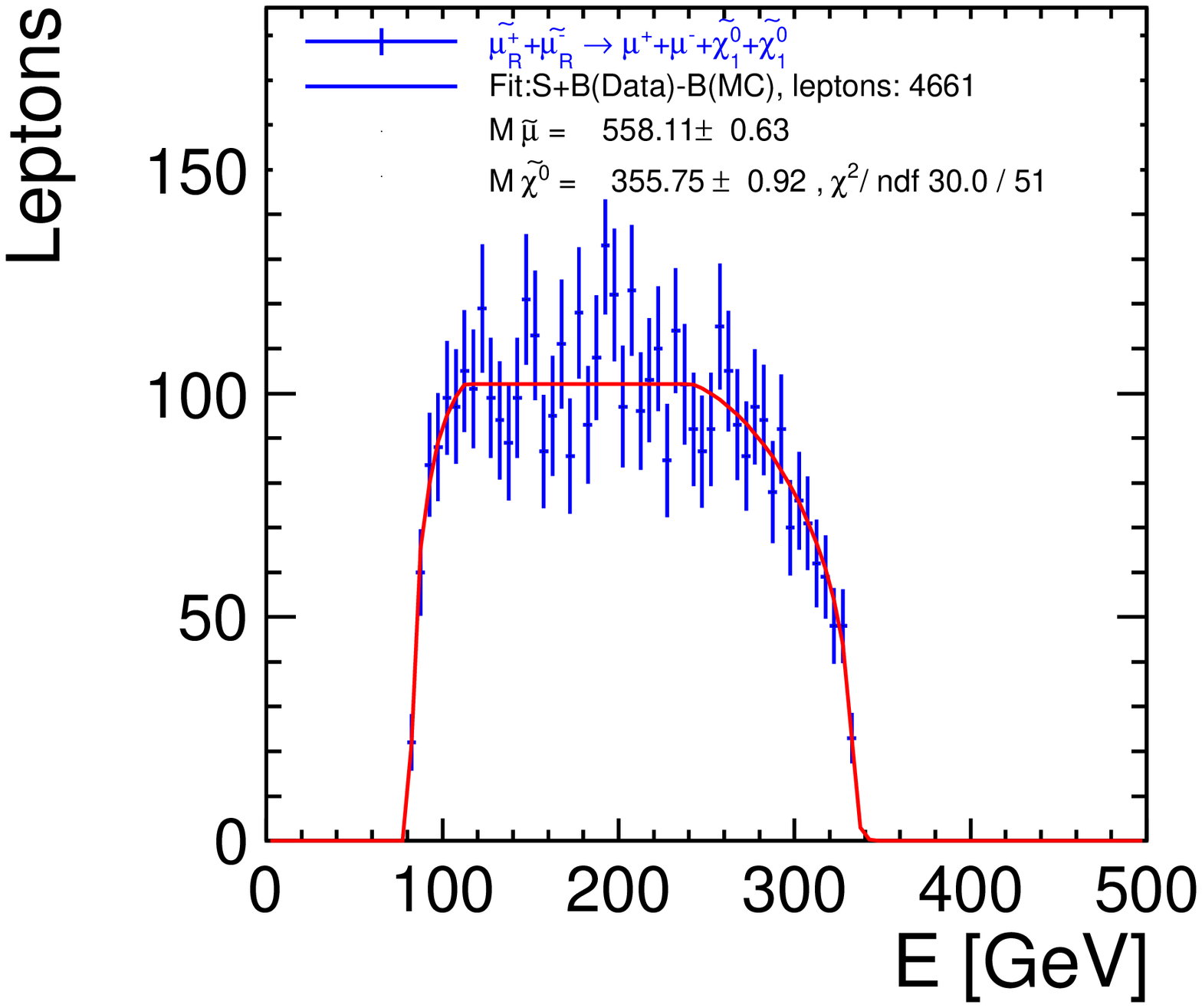}}
\subfloat[Events selected with a classifier trained with the true signal masses.]
{\includegraphics[width=0.50\textwidth]
{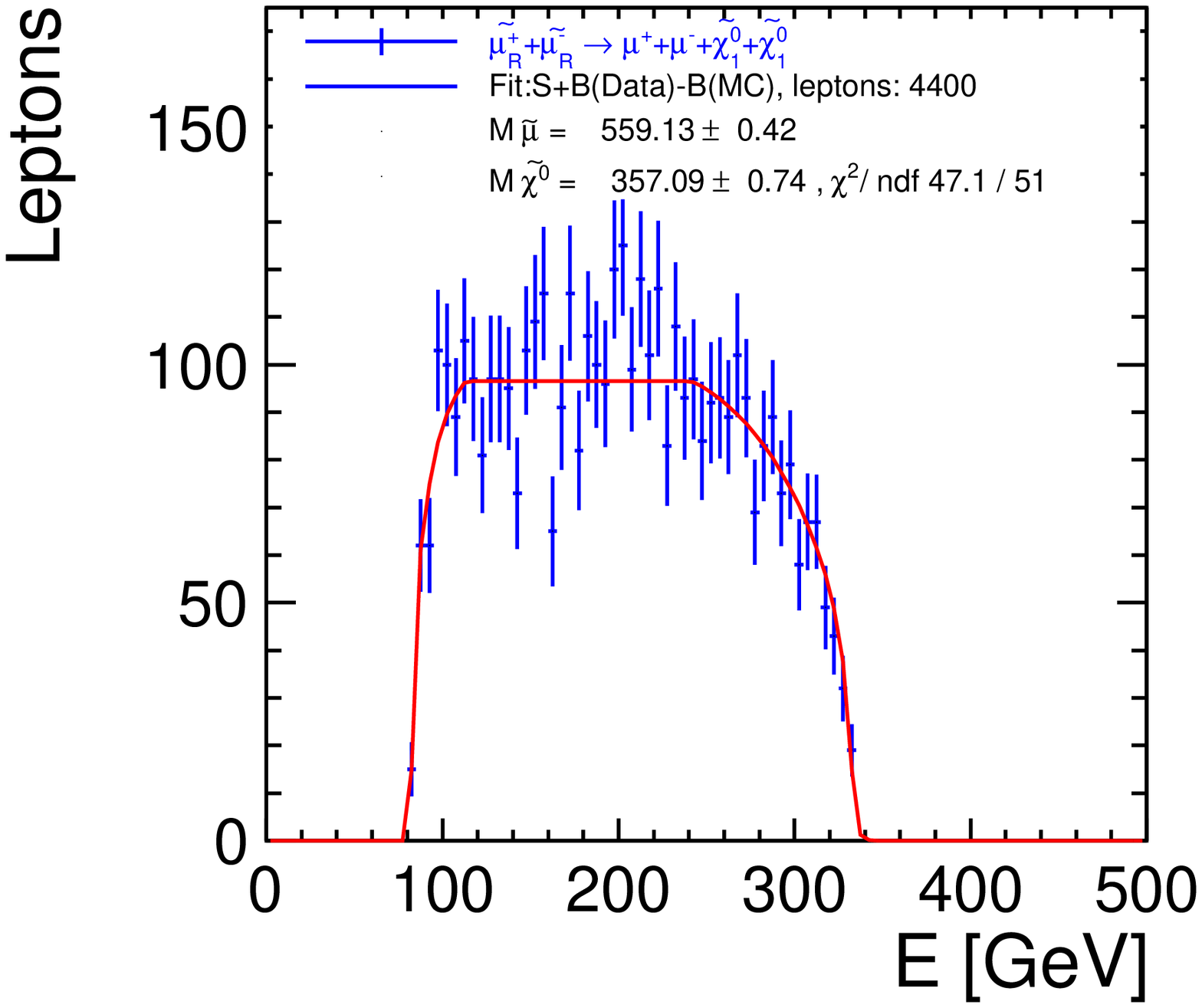}}
\end{tabular}
}  
\caption{Process $\ee \rightarrow  \tilde \mu_R^+ \tilde \mu_R^- $ at $\sqrt{s}=$ \mbox{1.4 TeV}:
muon energy distribution and fit results for events selected with a 9 variables classifier trained with signal masses lower by 10 GeV (a),
for events selected with a 9 variables classifier trained with the true signal masses (b).
}
\label{fig:205_mT-10}
\end{figure}

The signal energy distribution of the sample trained with larger masses is obviously biased, nevertheless
the end points are visible and have similar values as the ones of the sample trained with lower masses.
A rough estimation leads to $E_L$=80 GeV and $E_H$=340 GeV; with these values and $\sqrt(s)$=1.4 TeV formula 5.1
leads to a $\tilde \mu^{\pm}$ mass of 549.8 Gev and a $\neutralino{1}$ mass of 347.7 GeV,
these values are about 10 GeV lower than the true mass values.
These mass values are then used to simulate and reconstruct a new signal sample which is used to train 
the classifier with all 9 variables.
Figure~\ref{fig:205_mT-10} a) shows the the 
energy distribution of the data sample selected with the 9 variables classifier trained with masses lower by 10 GeV
and the fit result.
Figure~\ref{fig:205_mT-10} b) shows the the 
energy distribution of the data sample selected with a the 9 variables classifier trained with the true masses
and the fit result.
The smuon and neutralino masses determined after selection of the events with the classifier trained with masses lower by 10 GeV
are lower by 1.0 GeV and 1.3 GeV respectively; the values are statistically compatible  

\section{Systematic Uncertainty related to the Luminosity Spectrum }
The beam energy is derived from the beam deflection measurement using high precision beam position monitors (BPM) pairs 
placed before and after the first dipole in the energy collimation section. This setup provides
a relative energy resolution better than 0.04\% ~\cite{CDR-Vol1:2012}; therefore the impact on the slepton and gaugino    
masses is considered as negligible.

In this section the systematic uncertainty on the slepton and gaugino masses,
related to uncertainties in the the knowledge of the luminosity spectrum, is investigated.
The assessment is done at 3.0 TeV where the beamstrahlung is largest.
As can be seen from equation (\ref{formula:fitfunc}),
the slepton and gaugino masses depend on the effective luminosity function
\mbox{$L_{Eff}(\sqrt{s},\overrightarrow{p})= L(\sqrt{s}, \overrightarrow{p}) \otimes ISR(\sqrt{s}) \otimes \sigma_{\tilde\ell^{\pm}}(\sqrt{s})$}. 
The details about the method used to reconstruct the luminosity spectrum $L(\sqrt{s}, \overrightarrow{p})$ using
Bhabha events are reported in ~\cite{Andre:2013}.
The luminosity spectrum is parametrized with a function 
$F(x_1,x_2, \overrightarrow{p})$ where $x_{1,2}=2 E_{1,2} / \sqrt{s}$; $E_{1,2}$ is the energy of the $e^+ e^-$ particles before ISR;
the vector $\overrightarrow{p}$ has 19 parameters. The model takes into account the longitudinal boost, the correlation 
between the two particle energies and accounts for asymmetric beams.
A fit of $F(x_1,x_2,\overrightarrow{p})$ to the Bhabha events using the energy and the acollinearity of the outgoing  
$e^+ e^-$ particles allows to determine the parameters $\overrightarrow{p}$ of the luminosity function and their errors.
The parameters were determined at 3 TeV, using $ 2.2\cdot 10^{6}$ events and taking into account the $e^+ e^-$
energy resolution.

To estimate the systematic error on the masses due to the luminosity spectrum,
the mass fit is performed 38 times. Prior to each fit the effective luminosity spectrum 
$L_{Eff}(\sqrt{s},\overrightarrow{p})$
is computed; one parameter $p_i$ is changed to $p_i + \tfrac{ \sigma_{p_{i}}} {2}$ or $p_i - \tfrac{ \sigma_{p_{i}}} {2}$
and all other parameters are kept to their nominal value.

\noindent The error on the mass from the luminosity is:
\begin{eqnarray}
\sigma_m= \sqrt{\sum_{i,j}\delta_i C_{ij} \delta_j }
\end{eqnarray}
$C_{ij}$ is the correlation matrix obtained from the luminosity spectrum fit and:
\begin{eqnarray}
\delta_i = m(L_{Eff}(\sqrt{s}, \overrightarrow{p} +\overrightarrow{e}_{i} \frac{ \sigma_{p_i} } {2}))
-m(L_{Eff}(\sqrt{s},\overrightarrow{p} -\overrightarrow{e}_{i} \frac{\sigma_{p_i}}{2})).
\end{eqnarray}
where \textit{m} is the the result of the mass fit described in section 5.

Table~\ref{tab:syste3tev} shows the values of the slepton and gaugino masses, the corresponding statistical 
uncertainty, and the systematic errors from the knowledge of the shape of the luminosity spectrum.
For 2 $\mathrm{ab^{-1}}$ of integrated luminosity, the statistical errors are dominant.

%
\begin{table} [htbp]
\centering
\resizebox{\textwidth}{!} {
\begin{tabular}{ l c c c c c c }
\hline
Process &$m_{\tilde \ell}$ &$\sigma_m$ & $\sigma_m$  &$m_{\neutralino{1}}$ or $m_{\chargino{\pm}}$ &$\sigma_m$ &$\sigma_m$   \\
        &GeV               &(stat)\% &(lumi)\%  &GeV &(stat)\% &(lumi)\%            \\ \hline
$\ee \rightarrow \tilde\mu_R^+ \tilde\mu_R^-$  &1011.5  &0.45 &0.02  &341.7 &2.8 &0.06  \\
$\ee \rightarrow  \tilde e_R^+ \tilde e_R^-    $  &1011.8  &0.15 &0.02  &341.2 &0.80 &0.06 \\
$\ee \rightarrow  \tilde \nu_e \tilde \nu_e    $  &1093.6  &0.19 &0.02 &643.6 &0.54 &0.03  \\
\hline
\end{tabular}
}
\caption{Slepton and gaugino masses, statistical and systematic uncertainties, from the knowledge of the shape
of the luminosity spectrum (lumi) , at  $\sqrt{s}=$ 3 TeV.
}
\label{tab:syste3tev}
\end{table}
%

%
\section{Polarization}
Beam polarization is very helpful in the
study of SUSY processes both to improve the signal-to-background ratio and as an
analyzer~\cite{MoortgatPick:2005cw}, in particular to establish
the chirality of the sleptons.
Table~\ref{tab:cross_polar} shows the signal cross sections for
different electron and positron beam polarization conditions.
Running with left polarized electron beam would establish the
chirality of the selectron which decays into two leptons and of the selectron and sneutrinos
which decay into two leptons and four jets.
Running with right polarized electron beam would increase the cross sections of the
$\tilde \ell_R$ processes and reduce some of the backgrounds.


%
\begin{table} [htbp]
\caption{Signal processes cross sections ($\sigma$), for different electron, positron beam
polarization conditions, at  \mbox{$\sqrt{s}=$ 3 TeV}. }
\label{tab:cross_polar}
\centering
\begin{tabular}{ l c c c c }
\hline
Process & $\ee \rightarrow  \tilde \mu_R^+ \tilde \mu_R^-$ 
& $\ee \rightarrow  \tilde e_R^+ \tilde e_R^- $
& $\ee \rightarrow  \tilde e_L^+ \tilde e_L^- $ 
& $\ee \rightarrow  \tilde \nu_e \tilde \nu_e $  \\
\hline
                             &  $\sigma(fb)$ &$\sigma(fb)$ &$\sigma(fb)$ &$\sigma(fb)$ \\
beam polarization            &               & & & \\
$e^-$ :none~,~$e^+:$none     &~~~0.72       &~~~6.05           &3.06            &~~13.76\\
$e^-$ :L80\%,~$e^+: $none     &~~~0.46      &~~~2.59           &4.78            &~~21.90\\
$e^-$ :R80\%,~$e^+: $none     &~~~0.98      &~~~9.51           &1.34            &~~~~5.62 \\
$e^-$ :R80\%,~$e^+: $L60\%   &~~~1.15       &~11.40           &1.14            &~~~~4.56  \\
\hline
\end{tabular}
\end{table}

\section{Summary}
The accuracy of the slepton and gaugino mass determination and of the process cross section measurement 
in pair produced
$\tilde e_R, \tilde e_L, \tilde \mu_R$, and $\tilde \nu_e$ processes
has been studied at CLIC with the CLIC\_ILD\_CDR detector model for
two specific SUSY benchmark scenarios at \mbox{$\sqrt{s}=$ 3 TeV and 1.4 TeV}.
The analysis is based on two lepton and two lepton plus four jet final states.

The electron and muon energy resolution and the boson mass resolution
are not affected by the beam induced background, provided the detectors have timing
capabilities of the order of 1 nsec allowing for the application of PFO selection cuts.
The reconstructed boson mass accuracy allows 
$W^{\pm}$ and light $H$ final states to be distinguished;
$b$ tagging improves the purity of the $W^{\pm}$ and $H$ samples.

Slepton cross sections, slepton and gaugino masses can be extracted from the
lepton energy distributions. 
At \mbox{3.0 TeV}, for 2.0 $\mathrm{ab^{-1}}$ of integrated luminosity
the relative statistical error on the masses 
is in the range of 0.15 to 0.45\% for the sleptons and in the range of 0.5 to 2.8\% for the gauginos.
At \mbox{1.4 TeV}, for 1.5 $\mathrm{ab^{-1}}$ of integrated luminosity, the relative statistical errors,
on the slepton and gaugino masses are in the range of 0.1 to 0.2\%.

A major source of smearing of the kinematic edges of the lepton energy spectrum
is beamstrahlung and ISR. 
The measurement of the luminosity spectrum with Bhabha events, allows a good
control of the beamstrahlung.
The systematic errors on the slepton and gauginos masses due to the knowledge of
the luminosity spectrum were estimated.
At \mbox{3.0 TeV} 
for 2.0 $\mathrm{ab^{-1}}$ of integrated luminosity the statistical errors
are larger than the systematic errors. 
%
%
\section{Acknowledgments}
We are grateful to the colleagues who contributed to this study.
Daniel~Schulte for making available the beam spectrum and
the $\gamma \gamma \to \mathrm{hadron}$ background events.
D.Schlatter and A.Lucaci-Timoce for a thorough reading of the manuscript.

%


\begin{thebibliography}{100}

\bibitem{LCWS11}
M. Battaglia, J-J Blaising, J. Marshall, J> Nardulli, M. Thomson, A. Sailer, E. van der Kraaij, 
``Physics performances for Scalar Electrons, Scalar Muons and Scalar Neutrinos searches at CLIC,'' arXiv:1201.2092 [hep-ex]

\bibitem{Whizard:2008}
W. Kilian, T. Ohl, J. Reuter, ``WHIZARD, Simulating Multi-Particle Processes at LHC and ILC,'' arXiv: 0708.4233 [hep-ph]

\bibitem{Sjostrand:2006za} 
  T.~Sjostrand, S.~Mrenna and P.~Z.~Skands, ``PYTHIA 6.4 Physics and Manual,''
  JHEP {\bf 0605} (2006) 026, [arXiv:hep-ph/0603175].

\bibitem{Alwall:2006yp}
J.~Alwall {\it et al.},  
``A standard format for Les Houches event files,'' Comput. Phys. Commun., 176, 300-304, 2007.

\bibitem{Braun:2008zzb}
H.~Braun {\it et al.}  [CLIC Study Team], ``CLIC 2008 PARAMETERS,'' CLIC-NOTE-764 (2008).

\bibitem{c:thesis}
  D.~Schulte, ``Study of Electromagnetic and Hadronic Background in the Interaction Region of the TESLA Collider,''
  TESLA Note 97-08.

\bibitem{Agostinelli:2002hh}          
  S.~Agostinelli {\it et al.}  [GEANT4 Collaboration],                 
  ``GEANT4: A simulation toolkit,''            
  Nucl.\ Instrum.\ Meth.\  A {\bf 506} (2003) 250.

\bibitem{MoradeFreitas:2004sq}
  P.~Mora de Freitas,
  ``Mokka, Main Guidelines And Future,''
  in Proc.\ of the {\it Int. Conf. on Linear Colliders (LCWS 04)}  vol.\ 1 (2004) 441.

\bibitem{geom:2011}
 A. M{\"u}nnich, A. Sailer,
 ``The CLIC\_ILD\_CDR Geometry for the CDR Monte Carlo Mass Production,''
 CERN LCD-Note-2011-002

\bibitem{loi:2009}
 T. Abe {\it et al.} 
 [ILD Concept Group - Linear Collider Collaboration]
 ``The International Large Detector: Letter of Intent, 2010,''
 arXiv:1006.3396 [hep-ex]. 

\bibitem{Gaede:2006pj}
  F.~Gaede, ``Marlin and LCCD: Software tools for the ILC,''
  Nucl.\ Instrum.\ Meth.\  A {\bf 559} (2006) 177.                

\bibitem{Marshall:2010}          
 M. A. Thomson, ``Particle Flow Calorimetry and the PandoraPFA Algorithm,''
 Nucl.\ Instrum.\ Meth.\ A {\bf 611} (2009) 25.

\bibitem{Alster:2011}          
 N. Alster, M. Battaglia,
 ``Determination of Chargino and Neutralino Masses in high-mass SUSY scenarios at CLIC,''
 CERN LCD-Note-2011-003, arXiv:1104.0523 [hep-ex].

\bibitem{LCD:2011-028}
 J.S. Marshall, A. M{\"u}nnich, M.A. Thomson, 
 ``Performance of particle flow calorimetry at CLIC,'' 
 Nucl.\ Instrum.\ Meth.\ A {\bf 700} (2013) 153.

\bibitem{LCD:2011-020}
 T. Barklow, D. Dannheim, M. O. Sahin, D. Schulte,
 ``Simulation of $\gamma \gamma$ to hadron background at CLIC,'' 
 CERN LCD-Note-2011-020


\bibitem{Fastjet:2010}
 M. Cacciari, G. P. Salam and G. Soyez, ``The anti-kt jet clustering algorithm,''
 JHEP 0804 (2008) 063 [arXiv:0802.1189].

\bibitem{LCD:2010-006}
 M. Battaglia, P. Ferrari,
 ``A Study of $\ee \to H^0A^0 \to b \bar b b \bar b$ at 3 TeV at CLIC,''
 CERN LCD-Note-2010-006, arXiv:1006.5659 [hep-ex].

\bibitem{overlay:2011}
 P. Schade, A. Lucaci-Timoce,
 ``Description of the signal and background event mixing as implemented in the Marlin processor OverlayTiming,''
 CERN LCD-Note-2011-006

\bibitem{TMVA:2007}
   A.~Hoecker, P.~Speckmayer, J.~Stelzer, J.~Therhaag, E.~von Toerne, and H.~Voss,
   ``TMVA: Toolkit for Multivariate Data Analysis,''
   PoS A CAT 040 (2007) [physics/0703039].

\bibitem{Feng:1993sd}
  J.~L.~Feng and D.~E.~Finnell,
  ``Squark mass determination at the next generation of linear $e^+ e^-$ colliders,''
  Phys.\ Rev.\  D {\bf 49} (1994) 2369
  [arXiv:hep-ph/9310211].

\bibitem{Martyn:1999tc}
  H.~U.~Martyn and G.~A.~Blair,
  ``Determination of sparticle masses and SUSY parameters,''
  arXiv:hep-ph/9910416.

\bibitem{James:1975dr}
  F.~James and M.~Roos,
  ``Minuit: A System For Function Minimization and Analysis of the Parameter Errors And Correlations,''
  Comput.\ Phys.\ Commun.\  {\bf 10} (1975) 343.




\bibitem{CDR-Vol1:2012}
 ``A MULTI-TEV LINEAR COLLIDER BASED ON CLIC TECHNOLOGY, CLIC CONCEPTUAL DESIGN REPORT'',
 CERN 2012-007

\bibitem{Andre:2013}
 A, Sailer, S. Poss,
 ``Differential Luminosity Measurement using Bhabha Events,''
 CERN LCD-Note-2011-040

\bibitem{MoortgatPick:2005cw}
  G.~A.~Moortgat-Pick {\it et al.},
  ``The role of polarized positrons and electrons in revealing fundamental interactions at the linear collider,''
  Phys.\ Rept.\  {\bf 460} (2008) 131
  [arXiv:hep-ph/0507011].

\end{thebibliography}
\end{document}